\documentclass{article}
\usepackage[utf8]{inputenc}
\usepackage[margin=1in]{geometry}
\usepackage{tcs-macros}
\usepackage{dsfont}
\newcommand{\ACB}{\mathrm{AffCB}}
\newcommand{\CB}{\mathrm{CB}}
\newcommand{\nmExt}{\mathrm{nmExt}}
\newcommand{\Ext}{\mathrm{Ext}}
\newcommand{\AffExt}{\mathrm{AffExt}}
\newcommand{\LExt}{\mathrm{LExt}}
\newcommand{\Prefix}{\mathrm{Prefix}}

\newcommand{\Samp}{\mathrm{Samp}}
\newcommand{\SD}[1]{\Delta\left(#1\right)}
\newcommand{\Reduce}{\mathrm{Reduce}}

\newcommand{\BFExt}{\mathrm{BFExt}}
\newcommand{\sfp}{\mathsf{p}}
\newcommand{\sfm}{\mathsf{m}}
\newcommand{\sfr}{\mathsf{r}}
\newcommand{\hmin}{\mathrm{H}_{\infty}}
\newcommand{\avghmin}{\widetilde{\mathrm{H}}_{\infty}}
\newcommand{\lra}{\leftrightarrow}
\newcommand{\Maj}{\mathrm{Maj}}
\newcommand{\Spec}{\mathrm{Spec}}
\newcommand{\linspan}{\mathrm{span}}
\newtheorem{THM}{Theorem}
\usepackage{subfiles}

\title{\textbf{Extractors for  Sum of Two Sources}}
\author{Eshan Chattopadhyay\thanks{Supported by NSF CAREER award 2045576} \\
Cornell University\\
\texttt{eshan@cs.cornell.edu}
 \and
Jyun-Jie Liao\footnotemark[1]\\
Cornell University\\
\texttt{jjliao@cs.cornell.edu}
}

\begin{document}

\maketitle
\begin{abstract}
We consider the problem of extracting randomness from \textit{sumset sources}, a general class of weak sources introduced by Chattopadhyay and Li (STOC, 2016). An $(n,k,C)$-sumset source $\mathbf{X}$ is a distribution on $\{0,1\}^n$ of the form $\mathbf{X}_1 + \mathbf{X}_2 + \ldots + \mathbf{X}_C$, where $\mathbf{X}_i$'s are independent sources on $n$ bits with min-entropy at least $k$. Prior extractors either required the number of sources $C$ to be a large constant or the min-entropy $k$ to be at least $0.51 n$. 

As our main result, we construct an explicit extractor for sumset sources in the setting of $C=2$ for min-entropy $\mathrm{poly}(\log n)$ and polynomially small error. We can further improve the min-entropy requirement  to $(\log n) \cdot (\log \log n)^{1 + o(1)}$ at the expense of worse error parameter of our extractor. We find applications of our sumset extractor for extracting randomness from other well-studied models of weak sources such as affine sources, small-space sources,  and interleaved sources.

Interestingly, it is unknown if a random function is an extractor for sumset sources.  We use techniques from additive combinatorics to show that it is a disperser, and further prove that an affine extractor works for an interesting subclass of sumset sources which informally corresponds to the ``low doubling" case
(i.e., the support of $\bX_1 + \bX_2$ is not much larger than $2^k$).

\end{abstract}
\section{Introduction}
Randomness is a powerful resource in computer since, and has been widely used in areas such as algorithm design, cryptography, distributed computing, etc. Most of the applications assume the access to perfect randomness, i.e. a stream of uniform and independent random bits. However, natural sources of randomness often generate biased and correlated random bits, and in cryptographic applications there are many scenarios where the adversary learns some information about the random bits we use. This motivates the area of randomness extraction, which aims to construct randomness extractors which are deterministic algorithms that can convert an imperfect random source into a uniform random string. 

Formally, the amount of randomness in an imperfect random source $\bX$ is captured by its \emph{min-entropy}, which is defined as $\hmin(\bX)=\min_{x\in\Supp(\bX)}(-\log(\pr{\bX=x}))$.\footnote{$\Supp(\bX)$ denotes the support of $\bX$. We use $\log$ to denote the base-$2$ logarithm in the rest of this paper.} We call $\bX\in\bits{n}$ a $(n,k)$-source if it satisfies $\hmin(\bX)\ge k$. Ideally we want a deterministic function $\Ext$ with entropy requirement $k\ll n$, i.e. for every $(n,k)$-source $\bX$ the output $\Ext(\bX)$ is close to a uniform string. Unfortunately, a folklore result shows that it is impossible to construct such a function even when $k=n-1$. 

To bypass the impossibility result, researchers have explored two different approaches. The first one is based on the notion of \emph{seeded extraction}, introduced by Nisan and Zuckerman~\cite{NZ96}. This approach assumes that the extractor has access to a short independent uniform random seed, and the extractor needs to convert the given source $\bX$ into a uniform string with high probability over the seed. Through a successful line of research we now have seeded extractors with almost optimal parameters~\cite{LRVW03,GUV09,DKSS13}. In this paper, we focus on the second approach, called \emph{deterministic extraction}, which assumes some structure in the given source. Formally, a deterministic extractor is defined as follows.
\begin{definition}
Let $\cX$ be a family of distribution over $\bits{n}$. We say a deterministic function $\Ext:\bits{n}\to\bits{m}$ is a deterministic extractor for $\cX$ with error $\eps$ if for every distribution $\bX\in\cX$, 
$$\Ext(\bX)\approx_{\eps} \bU_m.$$
We say $\Ext$ is explicit if $\Ext$ is computable by a polynomial-time algorithm.
\end{definition}

The most well-studied deterministic extractors are multi-source extractors, which assume that the extractor is given $C$ independent $(n,k)$-sources $\bX_1,\bX_2,\dots,\bX_C$. This model was first introduced by Chor and Goldreich~\cite{CG88}. They constructed explicit two-source extractors with error $2^{-\Omega(n)}$ for entropy $0.51n$, and proved that there exists a two-source extractor for entropy $k=O(\log(n))$ with error $2^{-\Omega(k)}$. Significant progress was made by Chattopadhyay and Zuckerman~\cite{CZ19}, who showed how to construct an extractor for two sources with entropy $k=\polylog(n)$, after a long line of successful work on independent source extractors (see the references in \cite{CZ19}). The output length was later improved to $\Omega(k)$ by Li~\cite{Li16}. Furthermore, Ben-Aroya, Doron and Ta-Shma~\cite{BDT17} showed how to improve the entropy requirement to $O(\log^{1+o(1)}(n))$ for constant error and $1$-bit output. The entropy requirement was further improved in subsequent works~\cite{Coh17,Li17}, and the state-of-the-art result is by Li~\cite{Li19}, which requires $k=O(\log(n)\cdot \frac{\log\log(n)}{\log\log\log(n)})$. For a more elaborate discussion, see the survey by Chattopadhyay~\cite{Cha20survey}.

Apart from independent sources, many other classes of sources have been studied for deterministic extraction. We briefly introduce some examples here. A well-studied class is oblivious bit-fixing sources~\cite{CGH+85,GRs06,KZ07,Rao09}, which is obtained by fixing some bits in a uniform random string. Extractors for such sources have found applications in cryptography~\cite{CGH+85,KZ07}. A natural generalization of bit-fixing sources is the class of affine sources, which are uniform distributions over some affine subspaces and have been widely studied in literature (see \cite{CGL21} and references therein). Another important line of work focuses on the class of samplable sources, which are sources sampled by a ``simple procedure" such as efficient algorithms~\cite{TV00}, small-space algorithms~\cite{KRVZ11} or simple circuits~\cite{Vio14}. Researchers have also studied interleaved sources~\cite{RY11,CZ16-interleaved,CL16-sumset,CL20}, which is a generalization of independent sources such that the bits from different independent sources are permuted in an unknown order. 

In this paper, we consider a very general class of sources called \emph{sumset sources}, which was first studied by Chattopadhyay and Li~\cite{CL16-sumset}. A sumset source is the sum (XOR) of multiple independent sources, which we formally define as follows.
\begin{definition}
A source $\bX$ is a \emph{$(n,k,C)$-sumset source} if there exist $C$ independent $(n,k)$-sources $\{\bX_i\}_{i\in[C]}$ such that $\bX=\sum_{i=1}^C \bX_i$.
\end{definition}
Chattopadhyay and Li~\cite{CL16-sumset} showed that the class of sumset sources generalize many different classes we mentioned above, including oblivious bit-fixing sources, independent sources, affine sources and small-space sources. They also constructed an explicit extractor for $(n,k,C)$-sumset sources where $k=\polylog(n)$ and $C$ is a large enough constant, and then used the extractor to obtain new extraction results for small-space sources and interleaved $C$ sources. An interesting open question left in \cite{CL16-sumset} is whether it is possible to construct an extractor for $(n,\polylog(n),2)$-sumset source. An explicit construction of such an extractor would imply improved results on extractors for interleaved sources and small-space sources with polylogarithmic entropy. (We discuss the details in \Cref{sec:intro:applications}.)

However, it has been challenging to construct such an extractor for low min-entropy. The only known extractor for sum of two sources before this work is the Paley graph extractor~\cite{CG88}, which requires one source to have entropy $0.51n$ and the other to have entropy $O(\log(n))$, based on character sum estimate by Karatsuba~\cite{Kar71,Kar91} (see also \cite[Theorem 4.2]{CZ16-interleaved}). In fact, unlike other sources we mentioned above, it is not  clear whether a random function is an extractor for sumset sources. (See \Cref{sec:intro:random-function} for more discussions.)

In this paper, we give a positive answer to the question above. Formally, we prove the following theorem.
\begin{THM}\label{thm:ext-poly-error}
There exists a universal constant $C$ such that for every $k\ge \log^C(n)$, there exists an explicit extractor $\Ext:\bits{n}\to\bits{m}$ for $(n,k,2)$-sumset source with error $n^{-\Omega(1)}$ and output length $m=k^{\Omega(1)}$.
\end{THM}
We can further lower the entropy requirement to almost logarithmic at the expense of worse error parameter of the extractor. 
\begin{THM}\label{thm:ext-const-error}
For every constant $\eps>0$, there exists a constant $C_{\eps}$ such that there exists an explicit extractor $\Ext:\bits{n}\to\bits{}$ with error $\eps$ for $(n,k,2)$-sumset source where $k=C_{\eps}\log(n)\log\log(n)\log\log\log^3(n)$.
\end{THM}
Since a sumset source extractor is also an affine extractor, \Cref{thm:ext-const-error} also gives an affine extractor with entropy $O(\log(n)\log\log(n)\log\log\log^3(n))$, which slightly improves upon the $O(\log(n)\log\log(n)\log\log\log^6(n))$ bound in \cite{CGL21}. We note that this improvement comes from a new construction of ``affine correlation breakers", which we discuss in \Cref{sec:intro:acb}.

\subsection{Applications}\label{sec:intro:applications}
Next we show applications of our extractors to get improved extractors for other well-studied models of weak sources.
\subsubsection{Extractors for Interleaved Sources}
Interleaved sources are a natural generalization of two independent sources, first introduced by Raz and Yehudayoff~\cite{RY11} with the name ``mixed-2-sources". The formal definition of interleaved sources is as follows. For a $n$-bit string $w$ and a permutation $\sigma:[n]\to{[n]}$, we use $w_\sigma$ to denote the string such that the $\sigma(i)$-th bit of $w_\sigma$ is exactly the $i$-th bit of $w$. For two strings $x,y$ we use $x\circ y$ to denote the concatenation of $x$ and $y$. 
\begin{definition}
Let $\bX_1$ be a $(n,k_1)$-source, $\bX_2$ be a $(n,k_2)$-source independent of $\bX_1$ and $\sigma:[2n]\to{[2n]}$ be a permutation. Then $(\bX_1 \circ \bX_2)_\sigma$ is a \emph{$(n,k_1,k_2)$-interleaved sources}, or a \emph{$(n,k_1)$-interleaved sources} if $k_1=k_2$.
\end{definition}
Such sources naturally arise in a scenario that the bits of the input source come remotely from two independent sources in an unknown but fixed order. Furthermore, Raz and Yehudayoff~\cite{RY11} showed that an explicit extractor for such sources implies a lower bound for best-partition communication complexity.

Raz and Yehudayoff~\cite{RY11} constructed an extractor for $(n,(1-\beta)n)$-interleaved sources with $2^{-\Omega(n)}$ error for a small constant $\delta>0$. Subsequently, Chattopadhyay and Zuckerman~\cite{CZ16-interleaved} constructed an extractor for $(n,(1-\gamma)n,O(\log(n)))$-interleaved sources with error $n^{-\Omega(1)}$ for a small constant $\gamma>0$. A recent work by Chattopadhay and Li~\cite{CL20} gave an extractor for $(n,(2/3+\delta)n)$-interleaved sources with error $2^{-n^{\Omega(1)}}$, where $\delta$ is an arbitrarily small constant. In summary, all prior works required at least one of the sources to have min-entropy at least $0.66n$.

Observe that interleaved sources is a special case of sumset sources, as $(\bX_1\circ \bX_2)_\sigma=(\bX_1\circ 0^n)_\sigma + (0^n\circ \bX_2)_\sigma$. With our extractors for sum of two sources, we obtain the first extractors for interleaved two sources with polylogarithmic entropy.
\begin{corollary}
There exists a universal constant $C$ such that for every $k\ge \log^C(n)$, there exists an explicit extractor $\Ext:\bits{n}\to\bits{m}$ for $(n,k)$-interleaved sources with error $n^{-\Omega(1)}$.
\end{corollary}
\begin{corollary}
For every constant $\eps>0$, there exists a constant $C_{\eps}$ and an explicit extractor $\Ext:\bits{n}\to\bits{}$ with error $\eps$ for $(n,k)$-interleaved sources where $k=C_{\eps}\log(n)\log\log(n)\log\log\log^3(n)$.
\end{corollary}

We note that the above results easily extend to the setting when the two interleaved sources are of different lengths. In particular, this captures the following natural setting of ``somewhere independence": suppose we have a source $\bX$ on $n$ bits such that for some (unknown) $i$, the sources $\bX_{\le i}$ (first $i$ bits of $\bX$) and $\bX_{>i}$ (the last $n-i$ bits of $\bX$) are independent and each have entropy at least $k$. As long as $k \ge \poly(\log n)$, we can use our sumset extractor to extract from such sources. 

\subsubsection{Small-space Sources}
Kamp, Rao, Vadhan and Zuckerman~\cite{KRVZ11} first studied extractors for sources sampled by algorithms with limited memory. We define such small-space sources more formally as follows. 
\begin{definition}\label{def:small-space-source}
A space-$s$ sampling procedure $\cA$ with $n$-bit output is defined as follows. For every $(i,j)$ s.t. $i\in\bbZ,0\le i<n$ and $j\in\bits{s}$, let $\cD_{i,j}$ be a distribution over $\bits{}\times\bits{s}$. Then $\cA$ maintains an internal state $\mathsf{state}\in\bits{s}$, which is initially $0^s$, and runs the following steps for time step $i$ from $0$ to $n-1$: 
\begin{enumerate}
\item 
Sample $(x_{i+1},\mathsf{nextstate})\in\bits{}\times\bits{s}$ from $\cD_{i,\mathsf{state}}$.
\item
Output $x_{i+1}$, and assign $\mathsf{state}:=\mathsf{nextstate}$.
\end{enumerate}
Furthermore, the distribution $\bX$ of the output $(x_1,\dots,x_n)$ is called a space-$s$ source. 
\end{definition}
Equivalently, a space-$s$ source is sampled by a ``branching program" of width $2^s$ (see \Cref{sec:branching-program} for the formal definition). In \cite{KRVZ11} they constructed an extractor for space-$s$ source with entropy $k\ge Cn^{1-\gamma} s^{\gamma}$ with error $2^{-n^{\Omega(1)}}$, for a large enough constant $C$ and a small constant $\gamma>0$. Chattopadhyay and Li~\cite{CL16-sumset} then constructed an extractor with error $n^{-\Omega(1)}$ for space-$s$ source with entropy $k\ge s^{1.1}2^{\log^{0.51}(n)}$ based on their sumset source extractors. Recently, based on a new reduction to affine extractors, Chattopadhyay and Goodman~\cite{CG21} improved the entropy requirement to $k\ge s\cdot \polylog(n)$ (or $k\ge s\log^{2+o(1)}(n)$ if we are only interested in constant error and one-bit output).\footnote{Here we focus on the small-space extractors which minimize the entropy requirement. For small-space extractors with negligible error, see \cite{CG21} for a survey.} 

With our new extractors for sum of two sources and the reduction in \cite{CL16-sumset}, we can get extractors for space-$s$ source with entropy $s\log(n)+\polylog(n)$, which is already an improvement over the result in \cite{CG21}. In this work we further improve the reduction and obtain the following theorems. 
\begin{THM}\label{thm:small-space-ext-poly-error}
There exists a universal constant $C$ such that for every $s$ and $k\ge 2s+\log^C(n)$, there exists an explicit extractor $\Ext:\bits{n}\to\bits{m}$ with error $n^{-\Omega(1)}$ and output length $m=(k-2s)^{\Omega(1)}$ for space-$s$ sources with entropy $k$.
\end{THM}
\begin{THM}\label{thm:small-space-ext-const-error}
For every constant $\eps>0$, there exists a constant $C_{\eps}$ such that there exists an explicit extractor $\Ext:\bits{n}\to\bits{}$ with error $\eps$ for space-$s$ sources with entropy $2s+C_\eps\log(n)\log\log(n)\log\log\log^3(n)$.
\end{THM}
Interestingly, the entropy requirement of our extractors has \emph{optimal} dependence on the space $s$, since \cite{KRVZ11} showed that it is impossible to construct an extractor for space-$s$ source with entropy $\le 2s$. Moreover, the entropy in \Cref{thm:small-space-ext-const-error} almost matches the non-constructive extractor in \cite{KRVZ11} which requires entropy $2s+O(\log(n))$. 

\subsection{Affine Correlation Breakers}\label{sec:intro:acb}
One of the important building blocks of our sumset source extractors is an affine correlation breaker. While such an object has been constructed in previous works~\cite{Li16,CL16-sumset,CGL21}, in this paper we give a new construction with slightly better parameters. The main benefit of our new construction is that it is a \emph{black-box reduction} from affine correlation breakers to (standard) correlation breakers, which are simpler and more well-studied. We believe this result is of independent interest.

First we define a (standard) correlation breaker. Roughly speaking, a correlation breaker takes a source $\bX$ and a uniform seed $\bY$, while an adversary controls a ``tampered source" $\bX'$ correlated with $\bX$ and a ``tampered seed" $\bY'$ correlated with $\bY$. The goal of the correlation breaker is to ``break the correlation" between $(\bX,\bY)$ and $(\bX',\bY')$, with the help of some ``advice" $\alpha,\alpha'$. One can also consider the ``multi-tampering" variant where there are many tampered sources and seeds, but our theorem only uses the single-tampering version which is defined as follows.
\begin{definition}
$\CB:\bits{n}\times\bits{d}\times\bits{a}\to\bits{m}$ is a correlation breaker for entropy $k$ with error $\eps$ (or a $(k,\eps)$-correlation breaker for short) if for every $\bX,\bX'\in\bits{n}$, $\bY,\bY'\in\bits{d}$, $\alpha,\alpha'\in\bits{a}$ such that
\begin{itemize}
\item 
$\bX$ is a $(n,k)$ source and $\bY$ is uniform
\item
$(\bX,\bX')$ is independent of $(\bY,\bY')$
\item
$\alpha\neq\alpha' $,
\end{itemize}
it holds that 
$$\left(\CB(\bX,\bY,\alpha),\CB(\bX,\bY',\alpha')\right)\approx_{\eps}\left(\bU_m,\CB(\bX,\bY',\alpha')\right).$$
\end{definition}
The first correlation breaker was constructed implicitly by Li~\cite{Li13} as an important building block of his independent-source extractor. Cohen~\cite{Coh16-flipflop} then formally defined and strengthened this object, and showed other interesting applications. Chattophyay, Goyal and Li~\cite{CGL16} then used this object to construct the first non-malleable extractor with polylogarithmic entropy, which became a key ingredient for the two-source extractor in \cite{CZ19}. Correlation breakers have received a lot of attention and many new techniques were introduced to improve the construction~\cite{Coh16-acb,CS16,CL16-NIPM,Coh16-advice,Coh17,Li17,Li19}. 

Affine correlation breakers were first introduced by Li in his construction of affine extractors~\cite{Li16}, and were later used in \cite{CL16-sumset} to construct sumset source extractors. An affine correlation breaker is similar to a (standard) correlation breaker, with the main difference being that it allows $\bX$ and $\bY$ to have an ``affine" correlation, i.e. $\bX$ can be written as $\bA+\bB$ where $\bA$ is independent of $\bY$ and $\bB$ is correlated with $\bY$. The formal definition is as follows.
\begin{definition}\label{def:strong-ACB-informal}
$\ACB:\bits{n}\times\bits{d}\times\bits{a}\to\bits{m}$ is a $t$-affine correlation breaker for entropy $k$ with error $\eps$ (or a $(t,k,\eps)$-affine correlation breaker for short) if for every distributions $\bX,\bA,\bB\in\bits{n}$, $\bY,\bY^{1},\dots,\bY^{t}\in\bits{d}$ and strings $\alpha,\alpha^{1},\dots,\alpha^t\in\bits{a}$ such that
\begin{itemize}
\item
$\bX=\bA+\bB$
\item 
$\hmin(\bA)\ge k$ and $\bY$ is uniform
\item
$\bA$ is independent of $(\bB,\bY,\bY^{1},\dots,\bY^{[t]})$
\item
$\forall i\in[t]$, $\alpha\neq \alpha^{i}$,
\end{itemize}
it holds that 
$$\left(\ACB(\bX,\bY,\alpha),\{\ACB(\bX,\bY^{i},\alpha^{i})\}_{i\in[t]}\right)\approx_{\gamma}\left(\bU_m,\{\ACB(\bX,\bY^{i},\alpha^{i})\}_{i\in[t]}\right).$$
We say $\ACB$ is strong if 
$$\left(\ACB(\bX,\bY,\alpha),\bY,\{\ACB(\bX,\bY^{i},\alpha^{i}),\bY^i\}_{i\in[t]}\right)\approx_{\gamma}\left(\bU_m,\bY,\{\ACB(\bX,\bY^{i},\alpha^{i}),\bY^i\}_{i\in[t]}\right).$$
\end{definition}
The first affine correlation breaker in \cite{Li16} was constructed by adapting techniques from the correlation breaker construction in \cite{Li13} to the affine setting. Chattopadhyay, Goodman and Liao~\cite{CGL21} then constructed an affine correlation breaker with better parameters based on new techniques developed in more recent works on correlation breakers~\cite{Coh16-flipflop,CS16,CL16-NIPM,Li17}. 

While the techniques for standard correlation breakers can usually work for affine correlation breakers, it requires highly non-trivial modification, and it is not clear whether the ideas in the standard setting can always be adapted to the affine setting. In fact, the parameters of the affine correlation breaker in \cite{CGL21} do not match the parameters of the state-of-the-art standard correlation breaker by Li~\cite{Li19}, because adapting the ideas in \cite{Li19} to the affine setting (without loss in parameters) seems to be difficult. Moreover, it is likely that more improvements will be made in the easier setting of standard correlation breakers in the future, so a black-box reduction from affine correlation breakers to standard correlation breakers without loss in parameters will be very useful. In this work, we prove the following theorem.
\begin{THM}\label{thm:acb-to-std}
Let $C$ be a large enough constant. Suppose that there exists an explicit $(d_0,\eps)$-strong correlation breaker $\CB:\bits{d}\times\bits{d_0}\times\bits{a}\to\bits{C\log^2(t+1)\log(n/\eps)}$ for some $n,t\in\bbN$. Then there exists an explicit strong $t$-affine correlation breaker $\ACB:\bits{n}\times\bits{d}\times\bits{a}\to\bits{m}$ with error $O(t\eps)$ for entropy $k=O(td_0+tm+t^2\log(n/\eps))$, where $d=O(td_0+m+t\log^3(t+1)\log(n/\eps))$.
\end{THM}
As a corollary, by applying this black-box reduction on Li's correlation breaker~\cite{Li19}, we get an affine correlation breaker with parameters slightly better than those of \cite{CGL21}. (See \Cref{thm:acb} for more details.) As a result, our extractor in \Cref{thm:ext-const-error} only requires $O(\log(n)\log\log(n)\log\log\log^3(n))$ entropy, while using the affine correlation breaker in \cite{CGL21} would require $O(\log(n)\log\log(n)\log\log\log^6(n))$ entropy. 

In fact, if one can construct an ``optimal" standard correlation breaker with entropy and seed length $O(\log(n))$ when $t=O(1),a=O(\log(n)),\eps=n^{-\Omega(1)}$, which would imply a two-source extractor for entropy $O(\log(n))$, by \Cref{thm:acb-to-std} this also implies a sumset source extractor/affine extractor for entropy $O(\log(n))$.

\subsection{On Sumset Sources with Small Doubling}\label{sec:intro:random-function}
Finally we briefly discuss why a standard probabilistic method cannot be used to prove existence of extractors for sumset sources, and show some partial results about it. 

Suppose we want to extract from a source $\bA+\bB$, where $\bA$ and $\bB$ are independent $(n,k)$-sources. Without loss of generality we can assume that $\bA$ is uniform over a set $A$, and $\bB$ is uniform over another set $B$, such that $\abs{A}=\abs{B}=K$, where $K=2^k$. A simple calculation shows that there are at most $2^{2nK}$ choices of sources. In a standard probabilistic argument, we would like to show that a random function\footnote{A random function is sampled uniformly at random from all the possible choices of Boolean functions on $n$ input bits.} is an extractor for $\bA+\bB$ with probability at least $1-\delta$, where $\delta \ll 2^{-2nK}$, and then we could use union bound to show that a random function is an extractor for $(n,k,2)$-sources. However, this is not always true. For example, when $A=B$ is a linear subspace, then $\bA+\bB$ is exactly $\bA$, which has support size $K$. In this case we can only guarantee that a random function is an extractor for $\bA+\bB$ with probability $1-2^{-\beta K}$ for some $\beta<1$. In general, if the ``entropy" of $\bA+\bB$ is not greater than $k$ by too much, then the probabilistic argument above does not work. 
\begin{remark}
Note that the ``bad case" is not an uncommon case that can be neglected: if we take $A,B$ to be subsets of a linear space of dimension $k+1$, then $\abs{\Supp(\bA+\bB)}\le 2^{k+1}$, which means a random function is an extractor for $\bA+\bB$ with probability at most $1-2^{-2K}$. However, there are roughly $2^{4K}$ choices of $A$ and $B$, so even if we consider the bad cases separately the union bound still does not work.
\end{remark}

Nevertheless, we can use techniques from additive combinatorics to prove that the bad cases can be approximated with affine sources. With this result we can show that a random function is in fact a disperser\footnote{A disperser for a class of source $\cX$ is a boolean function $f$ which has non-constant output on the support of every $\bX\in \cX$.} for sumset sources. To formally define the bad cases, first we recall the definition of sumsets from additive combinatorics (cf. \cite{TV06}).
\begin{definition}
For $A,B\subseteq \bbF_2^n$, define $A+B=\{a+b:a\in A,b\in B\}$. For $A,B$ s.t. $\abs{A}=\abs{B}$ we say $(A,B)$ has doubling constant $r$ if $\abs{A+B}\le r\abs{A}$.
\end{definition}

It is not hard to see that a random function is a disperser for $\bA+\bB$ with probability exactly $1-2^{-\abs{A+B}+1}$. Therefore we can use union bound to show that a random function is a disperser with high probability for every sumset source $\bA+\bB$ which satisfies $\abs{A+B}> 3n\abs{A}$. When $\abs{A+B}\le 3n\abs{A}$, a celebrated result by Sanders~\cite{Sanders12} shows that $A+B$ must contain $90\%$ of an affine subspace with dimension $\log(\abs{A})-O(\log^4(n))$. With the well-known fact that a random function is an extractor for affine sources with entropy $O(\log(n))$, we can conclude that a random function is a disperser for sumset source with entropy $O(\log^4(n))$.

Note that Sanders' result only guarantees that $A+B$ almost covers a large affine subspace, but this affine subspace might only be a negligible fraction of $\bA+\bB$. Therefore, while a random function is an extractor for affine sources, Sanders' result only implies that it is a disperser for sumset source with small doubling constant. In this paper, we prove a ``distributional variant" of Sanders' result. That is, a sumset source $\bA+\bB$ with small doubling constant is actually statistically close to a convex combination of affine sources.
\begin{THM}\label{thm:affine-sumset}
Let $\bA,\bB$ be uniform distribution over $A,B\subseteq \bbF_2^n$ s.t. $\abs{A}=\abs{B}=2^k$ and $\abs{A+B}\le r\abs{A}$. Then $\bA+\bB$ is $\eps$-close to a convex combination of affine sources with entropy $k-O(\eps^{-2}\log (r)\log^3(r/\eps))$.
\end{THM}
Then we get the following corollary which says that an affine extractor is also an \emph{extractor} for sumset source with small doubling. 
\begin{corollary}\label{cor:random-function-sumset}
Let $\bA,\bB$ be uniform distribution over $A,B\subseteq \bbF_2^n$ s.t. $\abs{A}=\abs{B}=2^k$ and $\abs{A+B}\le r\abs{A}$. If $\mathrm{AffExt}:\bits{n}\to\bits{m}$ is an extractor for affine sources with entropy $k-\log^4(r)$, then $\AffExt(\bA+\bB)$ is $O(1)$-close to $\bU_m$. 
\end{corollary}
We remark that while \Cref{cor:random-function-sumset} implies that a random function is an extractor for sumset sources with small doubling, this does not mean a random function is an extractor for sumset sources in general. This is because a lower bound on $\abs{A+B}$ is not sufficient for us to show that a random function is an extractor by probabilistic argument. (See \Cref{appendix:random-function} for more discussions.)
\subsection{Open Problems}
In this paper we construct improved extractors for interleaved two sources and small-space sources based on our extractors for sum of two sources. Can we use our construction to get improved extractors for other classes of sources? More specifically, both of the applications require only an extractor for interleaved two sources, which is only a special case of sumset sources. Can we further exploit the generality of sumset sources?

Another interesting open problem is whether a random function is an extractor for sum of two sources. In this paper we prove that sumset sources have a ``structure vs randomness dichotomy": the sumset source is either close to an affine source, or has high enough entropy. In both cases a random function is a disperser. However our result does not seem strong enough to show that a random function is an extractor for sum of two sources. 

\section{Overview of Proofs}
In this section we give a high-level overview of our proofs. The overview includes some standard notations which can be found in \Cref{sec:prelim}.
\subsection{Construction of Sumset Extractors}
In this section we give an overview of construction of our sumset source extractors. Similar to \cite{CL16-sumset}, our extractor follows the two-step framework in \cite{CZ19}. First, we convert the sumset source into a non-oblivious bit-fixing (NOBF) source. Roughly speaking, a $t$-NOBF source is a string such that most of the bits are $t$-wise independent. (See \Cref{def:nobf} for the formal definition.) Second, we apply known extractors for NOBF sources~\cite{Vio14,CZ19,Li16,Meka17} to get the output. In the rest of this section, we focus on the first step, which is the main contribution of this work. 

\subsubsection{Reduction from Two Sources}
To see how our reduction works, first we recall the transformation from two independent sources to NOBF sources in \cite{CZ19}. Given two $(n,k)$-source $\bX_1,\bX_2$, first take a $t$-non-malleable extractor $\nmExt:\bits{n}\times\bits{d_1}\to\bits{}$ with error $\eps_1$, enumerate all the seeds and output a string $\bR_1:=\{\nmExt(\bX_1,s)\}_{s\in\bits{d_1}}$ with $D_1=2^{d_1}$ bits. We do not give the exact definition of non-malleable extractors here, but we need the following property proved in \cite{CZ19}: except for $\sqrt{\eps_1}$ fraction of ``bad bits", every $(t+1)$ ``good bits" in $\bR_1$ are $\sqrt{\eps_1}$-close to uniform. With this property it might seem like $\bR_1$ is close to a $(t+1)$-NOBF source, but unfortunately this is not true. While $\bR_1$ is guaranteed to be $D_1^{t+1}\sqrt{\eps_1}$-close to a NOBF source by a result in \cite{AGM03}, this bound is trivial since $D_1=\poly(1/\eps_1)$. To get around this problem, \cite{CZ19} used the second source $\bX_2$ to sample $D_2\ll D_1$ bits from $\bR_1$ and get $\bR_2$. Now $\bR_2$ is guaranteed to be $D_2^{t+1}\sqrt{\eps_1}$-close to a NOBF source, and the error bound $D_2^{t+1}\sqrt{\eps_1}$ can be very small since $D_2$ is decoupled from $\eps_1$. We note that Li~\cite{Li15} also showed a reduction from two independent sources to NOBF sources, and the sampling step is also crucial in Li's reduction.

Chattopadhyay and Li~\cite{CL16-sumset} conjectured that a similar construction should work for sumset sources. However, in the setting of sumset sources, it is not clear how to perform the sampling step. For example, if one replaces both $\bX_1$ and $\bX_2$ in the above construction with a sumset source $\bX=\bX_1+\bX_2$, then the sampling step might not work because the randomness we use for sampling is now correlated with $\bR_1$. Therefore, they adopted an idea in \cite{Li13} which requires the given source $\bX$ to be the sum of $C>2$ independent sources. In this paper, we show that we can actually make the sampling step work with a $(n,\polylog(n),2)$-sumset source. As a result we get an extractor for sum of two independent sources.

\subsubsection{Sampling with Sumset Source}
As a warm up, first we assume that we are sampling from the output of a ``$0$-non-malleable extractor", i.e. a strong seeded extractor. Let $\Ext:\bits{n}\times\bits{d_1}\to\bits{}$ be a strong seeded extractor with error $\eps_1$. First observe that the sampling method has the following equivalent interpretation. Note that $\Ext$ and the source $\bX_1$ together define a set of ``good seeds" such that a seed $s$ is good if $\Ext(\bX_1,s)$ is $\sqrt{\eps_1}$-close to uniform. Since $\Ext$ is a strong seeded extractor, $(1-\sqrt{\eps_1})$ of the seeds should be good. In the sampling step we apply a sampler $\Samp$ on $\bX_2$ to get some samples of seeds $\{\Samp(\bX_2,i)\}_{i\in\bits{d_2}}$. Then we can apply the function $\Ext(\bX_1,\cdot)$ on these sampled seeds to get the output $\bR_2=\{\Ext(\bX,\Samp(\bX,i))\}_{i\in\bits{d_2}}$ which is $2^{d_2}\sqrt{\eps_1}$-close to a $1$-NOBF source. 

Now we move to the setting of sumset sources and replace both $\bX_1,\bX_2$ in the above steps with $\bX=\bX_1+\bX_2$. Our goal is to show that we can still view this reduction as if we were sampling good seeds with $\bX_2$ and using these seeds to extract from $\bX_1$. Consider the $i$-th output bit, $\Ext(\bX,\Samp(\bX,i))$. Our main observation is, if $\Samp(\cdot,i)$ is a linear function, then we can assume that we compute $\Ext(\bX,\Samp(\bX,i))$ in the following steps:
\begin{enumerate}
\item 
First sample $x_2\sim \bX_2$.
\item
Use $x_2$ as the randomness of $\Samp$ to sample a ``seed" $s:=\Samp(\bX_2,i)$.
\item
Output $\Ext'_{x_2,i}(\bX_1,s):=\Ext(\bX_1+x_2,s+\Samp(\bX_1,i))$.
\end{enumerate}
First we claim that $\Ext'_{x_2,i}$ is also a strong seeded extractor. To see why this is true, observe that if we fix $\Samp(\bX_1,i)=\Delta$, then $\Ext'_{x_2,i}(\bX_1,\bU)=\Ext(\bX_1+x_2,\bU+\Delta)$. As long as $\bX_1$ still has enough entropy after fixing $\Samp(\bX_1,i)$, $\Ext$ works properly since $\bX_1+x_2$ is independent of $\bU+\Delta$, $\bX_1+x_2$ still has enough entropy and $\bU+\Delta$ is also uniform. Therefore, we can use $\Ext'_{x_2,i}$ and $\bX_1$ to define a set of good seeds $s$ which make $\Ext'_{x_2,i}(\bX_1,s)$ close to uniform, and most of the seeds should be good. Then we can equivalently view the sampling step as if we were sampling good seeds for $\Ext'_{x_2,i}$ using $\bX_2$ as the randomness. 

There are still two problems left. First, the definition of $\Ext'_{x_2,i}$ depends on $x_2$, which is the randomness we use for sampling. To solve this problem, we take $\Ext$ to be linear, and prove that $(1-\sqrt{\eps_1})$ fraction of the seeds $s$ are good in the sense that $\Ext'_{x_2,i}(\bX_1,s)$ is close to uniform for \emph{every} $x_2$. Second, $\Ext'_{x_2,i}$ depends on $i$, which is the index of our samples. Similarly we change the definition of good seeds so that a seed $s$ is good if $\Ext'_{x_2,i}(\bX_1,s)$ is good for every $x_2$ and $i$, and by union bound we can show that $(1-2^{d_2}\sqrt{\eps_1})$ fraction of the seeds are good. As long as $\eps_1\ll 2^{-2d_2}$, most of the seeds should be good. Now the definition of good seeds is decoupled from the sampling step, and hence we can show that most of the sampled seeds are good. 

\subsubsection{Sampling with Correlation Breakers}
Next we turn to the case of $t$-non-malleable extractors. Similar to how we changed the definition of good seeds for a strong seeded extractor, we need to generalize the definition of good seeds for a non-malleable extractor in \cite{CZ19} to the sumset source setting. First, we say a seed $s$ is good with respect to $x_2$ and a set of indices $T=\{i_1,\dots,i_{t+1}\}$ if for every $s^1,\dots,s^t\in\bits{d_1}$,
$$(\nmExt(\bX_1+x_2,s+\Samp(\bX_1,i_1))\approx_{\sqrt{\eps_1}}\bU_1) \mid \{\nmExt(\bX_1+x_2,s^j+\Samp(\bX_1,i_{j+1}))\}_{j\in[t]}.$$
Based on the proof in \cite{CZ19} and the arguments in the previous section, if $\bX_1$ has enough entropy when conditioned on $\{\Samp(\bX_1,i)\}_{i\in T}$, then $1-\sqrt{\eps_1}$ of the seeds are good with respect to $x_2$ and $T$. If we can prove that most of the seeds we sample using $x_2\sim \bX_2$ are good with respect to $x_2$ and every set of indices $T$, then the we can conclude that the output $\bR_2=\{\nmExt(\bX,\Samp(\bX,i))\}_{i\in\bits{d_2}}$ is $D_2^{t+1}\sqrt{\eps_1}$-close to a NOBF source. 

Next we need to show that most of the seeds are good with respect to \emph{every} $x_2$ and $T$, so that the sampling step is decoupled from the definition of good seeds. To deal with the dependence on $T$, we take the union bound over $T$, and we can still guarantee that $1-D_2^{t+1}\sqrt{\eps_1}$ of the seeds are good. To deal with the dependency on $x_2$, it suffices to replace the non-malleable extractor with a strong affine correlation breaker. Although the correlation breaker needs an additional advice string to work, here we can simply use the indices of the samples as the advice. Our final construction would be $\{\ACB(\bX,\Samp(\bX,\alpha),\alpha)\}_{\alpha\in\bits{d_2}}$.

Finally, we note that in order to make the extractor work for almost logarithmic entropy (\Cref{thm:ext-const-error}), we need to replace the sampler with a ``somewhere random sampler" based on the techniques in \cite{BDT17}, and the construction and analysis should be changed correspondingly. We present the details in \Cref{sec:sumset-ext}. 
\subsection{Reduction from Small-Space Sources to Sumset Sources}
In this section we give an overview of our new reduction from small-space sources to sumset sources. As in all the previous works on small-space source extractors, our reduction is based on a simple fact: conditioned on the event that the sampling procedure is in state $j$ at time $i$, the small-space source $\bX$ can be divided into two independent sources $\bX_1\in\bits{i},\bX_2\in\bits{n-i}$, such that $\bX_1$ contains the bits generated before time $i$, and $\bX_2$ contains the bits generated after time $i$. Kamp, Rao, Vadhan and Zuckerman~\cite{KRVZ11} proved that if we pick some equally distant time steps $i_1,\dots,i_{\ell-1}$ and condition on the states visited at these time steps, we can divide the small-space source into $\ell$ independent blocks such that some of them have enough entropy. However, such a reduction does not work for entropy smaller than $\sqrt{n}$ (cf. \cite{CG21}). Chattopadhyay and Li~\cite{CL16-sumset} observed that with a sumset source extractor we can extract from the concatenation of independent sources with \emph{unknown and uneven length}. They then showed that with a sumset source extractor, we can ``adaptively" pick which time steps to condition on and break the $\sqrt{n}$ barrier. Chattopadhyay and Goodman~\cite{CG21} further refined this reduction and showed how to improve the entropy requirement by reducing to a convex combination of affine sources. The reductions in \cite{CL16-sumset} and \cite{CG21} can be viewed as ``binary searching" the correct time steps to condition on, so that the given source $\bX$ becomes the concatenation of independent blocks $(\bX_1,\dots,\bX_{O(\log(n))})$ such that some of them have enough entropy. However, even though with our extractors for sum of two sources we only need two of the blocks to have enough entropy, the ``binary search-based" reduction would condition on at least $\log(n)$ time steps and waste $s\log(n)$ entropy. 

A possible way to improve this reduction is by directly choosing the ``correct" time step to condition on so that we only get two blocks $\bX_1\circ \bX_2$ both of which have enough entropy. However this is not always possible. For example, consider a distribution which is a convex combination of $\bU_{n/2}\circ 0^{n/2}$ and $0^{n/2}\circ \bU_{n/2}$. This distribution is a space-$1$ source and has entropy $n/2$, but no matter which time step we choose to condition on, one of the two blocks would have zero entropy. 

To resolve these problems, we carefully define the event to condition on as follows. For ease of explanation we view the space-$s$ sampling procedure as a branching program of width $2^s$. (Unfamiliar readers can consult \Cref{sec:branching-program}.) First, we define a vertex $v=(i,j)$ to be a ``stopping vertex" if the bits generated after visiting $v$ has entropy \emph{less} than some threshold. Then we condition on a random variable $\bV$ which is the \emph{first} stopping vertex visited by the sampling process. Note that $\bV$ is well-defined since every state at time $n$ is a stopping vertex. Besides, conditioning on $\bV$ only costs roughly $s+\log(n)$ entropy since there are only $n\cdot 2^s$ possible outcomes. 

Now observe that the event $\bV=(i,j)$ means the sampling process visits $(i,j)$ but does not visit any stopping vertex before time $i$. Let ``first block" denote the bits generated before time $i$ and ``second block" denote the bits generated after time $i$. It is not hard to see that the two blocks are still independent conditioned on $\bV=v$. Then observe that the first block has enough entropy because the second block does not contain too much entropy (by our definition of stopping vertex). Next we show that the second block also has enough entropy. For every vertex $u$, let $\bX_u$ denote the bits generated after visiting $u$. The main observation is, if there is an edge from a vertex $u$ to a vertex $v$, then unless $u\to v$ is a ``bad edge" which is taken by $u$ with probability $<\eps$, the entropy of $\bX_v$ can only be lower than $\bX_u$ by at most $\log(1/\eps)$. If we take $\eps\ll 2^{-s}n^{-1}$, then by union bound the probability that any bad edge is traversed in the sampling procedure is $\ll 1$. Since we take $\bV$ to be the \emph{first} vertex such that $\bX_\bV$ has entropy lower than some threshold, the entropy of $\bX_\bV$ can only be $\log(1/\eps)\approx s+\log(n)$ lower than the threshold. In conclusion, if we start with a space-$s$ source with entropy roughly $2s+2\log(n)+2k$, and pick the entropy threshold of the second block to be roughly $k+s+\log(n)$, we can get two blocks both having entropy at least $k$.

\subsection{From Affine to Standard Correlation Breaker}
Next we briefly discuss our black-box reduction from affine correlation breakers to standard correlation breakers. To reduce an affine correlation breaker to a standard correlation breaker, our main idea is similar to that of \cite{CGL21}: to adapt the construction of a correlation breaker from the independent-source setting to the affine setting, we only need to make sure that every function on $\bX$ is linear, and every function on $\bY$ works properly when $\bY$ is a weak source. However, instead of applying this idea step-by-step on existing constructions, we observe that every correlation breaker can be converted into a ``two-step" construction which is easily adaptable to the affine setting. First, we take a prefix of $\bY$ as the seed to extract a string $\bZ$ from $\bX$. Next, we apply a correlation breaker which treats $\bY$ as the source and $\bZ$ as the seed. This construction only computes one function on $\bX$, which is a seeded extractor and can be replaced with a linear one. Furthermore, the remaining step (i.e. the correlation breaker) is a function on $\bY$, which does not need to be linear. Finally, we note that if the underlying standard correlation breaker is strong, we can use the output as the seed to extract from $\bX$ linearly and get a strong affine correlation breaker.

A drawback of this simple reduction is that the resulting affine correlation breaker has a worse dependence on the number of tampering $t$. Recall that the state-of-the-art $t$-correlation breaker~\cite{Li19} requires entropy and seed length $O(t^2d)$ where $d=O\left(\log(n)\cdot \frac{\log\log(n)}{\log\log\log(n)}\right)$, assuming the error is $1/\poly(n)$ and the advice length is $\log(n)$. With the reduction above we get a $t$-affine correlation breaker with entropy and seed length $O(t^3d)$, while the affine correlation breaker in \cite{CGL21} has entropy and seed length $O(t^2\log(n)\log\log(n))$. Since the construction of sumset source extractors requires $t$ to be at least $\Omega(\log\log\log^2 (n))$, $O(t^3d)$ is actually worse than $O(t^2\log(n)\log\log(n))$. To improve the parameters, we first apply the reduction above to get a $1$-affine correlation breaker, and then strengthen the affine correlation breaker to make it work for $t$ tampering. Our strengthening procedure only consists of several rounds of alternating extractions, which requires $\poly(t)\cdot O(\log n)$ entropy. Therefore by plugging in the correlation breaker in \cite{Li19} we end up getting a $t$-affine correlation breaker with entropy and seed length $O(td+\poly(t)\cdot\log(n))$, which is better than $O(t^2\log(n)\log\log(n))$.

The strengthening procedure works as follows. Observe that the $1$-affine correlation breaker outputs a string $\bR$ which is uniform conditioned on \emph{every single} tampered version of $\bR$. (Note that $\bR$ might not be uniform when conditioned on all $t$ tampered versions simultaneously.) Then we apply alternating extractions to \emph{merge the independence of $\bR$ with itself}. Based on the ``independence merging lemma" in \cite{CGL21} (see \Cref{lemma:ind-merging}), after one round of alternating extraction, we get a string $\bR'$ which is uniform conditioned on \emph{every two} tampered $\bR'$. By repeating this step for $\log(t)$ times we get a $t$-affine correlation breaker.

\subsection{Sumset Sources with Small Doubling}
Finally we briefly sketch how to prove that a sumset source with small doubling is close to a convex combination of affine sources. Let $A,B\subseteq \bbF_2^n$ be sets of size $K=2^k$ and let $\bA,\bB$ be uniform distributions over $A,B$ respectively. A seminal result by Sanders~\cite{Sanders12} showed that there exists a large affine subspace $V$ such that at least $1-\eps$ fraction of $V$ is in $A+B$. We adapt Sanders' proof to show that for every distinguisher with output range $[0,1]$, the sumset source $\bA+\bB$ is indistinguishable from a convex combination of affine sources (with large entropy). Then by an application of von Neumann's minimax theorem (\Cref{coro:distribution-minimax}) we can find a universal convex combination of affine sources which is statistically close to $\bA+\bB$.

To see more details, first we briefly recall the outline of Sanders' proof. Consider $A',B'\subseteq\bbF_2^m$ such that $\abs{A'},\abs{B'}\ge \abs{\bbF_2^m}/r$, and let $\bA',\bB'$ be uniform distributions over $A',B'$ respectively. Let $\mathds{1}_{A'+B'}$ denote the indicator function for $A'+B'$. Based on the Croot-Sisask lemma~\cite{CS10} and Fourier analysis, Sanders showed that for arbitrarily small constant $\eps>0$ there exists a distribution $\bT\subseteq\bbF_2^m$ and a linear subspace $V$ of co-dimension $O(\log^4(r))$ s.t. $$\ex{\mathds{1}_{A'+B'}(\bA'+\bB')}\approx_{\eps}\ex{\mathds{1}_{A'+B'}(\bT+\bV)},$$
where $\bV$ is the uniform distribution over $V$. Then Sanders' original result follows directly by taking $\bT=t$ which maximizes $\ex{\mathds{1}_{A'+B'}(t+\bV)}$. 

A closer inspection at Sanders' proof shows that $\mathds{1}_{A'+B'}$ can be replaced with any function $f:\bbF_2^m\to{[0,1]}$. (Note that the distributions $\bT,\bV$ depend on the function $f$.) This implies that $\bA'+\bB'$ is indistinguishable from a convex combination of affine sources by $f$. With our minimax argument we can conclude that $\bA'+\bB'$ is statistically close to a convex combination of affine sources.

However, the result above only works for dense sets $A',B'$. To generalize the result to sets $A,B$ with small doubling, a standard trick in additive combinatorics is to consider a linear Freiman homomorphism $\phi:\bbF_2^n\to\bbF^m_2$, which is a linear injective function on $\ell A+\ell B$ for some constant $\ell$, and consider $A'=\phi(A),B'=\phi(B)$. By considering the function $f\circ\phi^{-1}$ we can still show that 
$$\ex{f(\bA+\bB)}= \ex{f(\phi^{-1}(\bA'+\bB'))}\approx\ex{f(\phi^{-1}(\bT+\bV))}.$$
However, it is not clear whether $\phi^{-1}(\bT+\bV)$ is a also a convex combination of affine sources in $\bbF_2^n$. To solve this problem, we adapt Sanders' proof to show that there exist $\bT,\bV$ which satisfy
\begin{equation}\label{eq:former}
\ex{\mathds{1}_{A'+B'}(\bA'+\bB')}\approx_{\eps}\ex{\mathds{1}_{A'+B'}(\bT+\bV)}
\end{equation}
and 
\begin{equation}\label{eq:later}
\ex{f(\phi^{-1}(\bA'+\bB'))}\approx_{\eps}\ex{f(\phi^{-1}(\bT+\bV))}
\end{equation}
\emph{simultaneously}. This relies on a variant of the Croot-Sisask lemma which shows that there exists a large set of ``common almost period" for $\mathds{1}_{A'+B'}$ and $f\circ \phi^{-1}$. Then (\ref{eq:former}) guarantees that with probability at least $1-2\eps$ over $t\sim \bT$, $\phi^{-1}(t+\bV)$ is an affine source in $\bbF_2^n$ with entropy $k-O(\log^4(r))$. Therefore $\phi^{-1}(\bT+\bV)$ is $2\eps$-close to a convex combination of affine sources. Finally (\ref{eq:later}) shows that $\bA+\bB$ is indistinguishable from $\phi^{-1}(\bT+\bV)$ by $f$, which implies our claim.

\paragraph{Organization.} In \Cref{sec:prelim} we introduce some necessary preliminaries and prior works. In \Cref{sec:small-space} we show a new reduction from small-space sources to sum of two sources which has optimal dependence on the space parameter, and prove \Cref{thm:small-space-ext-poly-error} and \Cref{thm:small-space-ext-const-error}. In \Cref{sec:sumset-ext} we show how to construct the extractors for sum of two sources in \Cref{thm:ext-poly-error} and \Cref{thm:ext-const-error}, assuming access to an affine correlation breaker. In \Cref{sec:affine-to-std} we show how to construct the affine correlation breaker we need based on a black-box reduction to a standard correlation breaker (\Cref{thm:acb-to-std}). Finally, we prove \Cref{thm:affine-sumset} in \Cref{sec:small-doubling}. 
\section{Preliminaries}\label{sec:prelim}
In this section we introduce some preliminaries. We note that \Cref{sec:branching-program} is only used in \Cref{sec:small-space}, \Cref{sec:seeded-extractors} to \ref{sec:ind-merging} are only used in \Cref{sec:sumset-ext} and \ref{sec:affine-to-std}, and \Cref{sec:additive-combinatorics} to \ref{sec:minimax} are only used in \Cref{sec:small-doubling}.
\subsection{Notations}
\paragraph{Basic notations.} The logarithm in this paper is always base 2. For every $n\in\bbN$, define $[n]=\{1,2,\dots,n\}$. In this paper, $\bits{n}$ and $\bbF_2^n$ are interchangeable, and so are $\bits{n}$ and $[2^n]$. We use $x\circ y$ to denote the concatenation of two strings $x$ and $y$. We say a function is explicit if it is computable by a polynomial time algorithm. For $x,y\in\bbR$ we use $x\approx_{\eps} y$ to denote $\abs{x-y}\le \eps$ and $x\not\approx_{\eps} y$ to denote $\abs{x-y}> \eps$. For every function $f:\cX\to \cY$ and set $A\subseteq \cX$, define $f(A)=\{f(x):x\in A\}$. For a set $A\subseteq \cX$ we use $\mathds{1}_A:\cX\to\bits{}$ to denote the indicator function of $A$ such that $\mathds{1}_A(x)=1$ if and only if $x\in A$. 

\paragraph{Distributions and random variables.} We sometimes abuse notation and treat distributions and random variables as the same. We always write a random variable/distribution in boldface font. We use $\Supp(\bX)$ to denote the support of a distribution. We use $\bU_n$ to denote the uniform distribution on $\bits{n}$. When $\bU_n$ appears with other random variables in the same joint distribution, $\bU_n$ is considered to be independent of other random variables. Sometimes we omit the subscript $n$ of $\bU_n$ if the length is less relevant and is clear in the context. When there is a sequence of random variables $\bX_1,\bX_2,\dots,\bX_t$ in the context, for every set $S\subseteq [t]$ we use $\bX_S$ to denote the sequence of random variables which use indices in $S$ as subscript, i.e. $\bX_S:=\{\bX_i\}_{i\in S}$. We also use similar notation for indices on superscript. 

\paragraph{Linear algebra.} For a set $A\subseteq \bbF_2^n$, we use $\linspan(A)$ to denote the linear span of $A$, and $A^\bot$ to denote the orthogonal complement of $\linspan(A)$, i.e. $A^{\bot}:=\{y\in\bbF_2^n: \forall x\in A, \inp{y,x}=0\}$. For every affine subspace $A$ of $\bbF_2^n$ we use $\dim(A)$ to denote the dimension of $A$. Note that if $\bA$ is uniform over $A$, then $\hmin(\bA)=\dim(A)$. Therefore we use ``dimension" and ``entropy" interchangeably when discussing affine sources.

\subsection{Statistical Distance}
\begin{definition}
Let $\bD_1,\bD_2$ be two distributions on the same universe $\Omega$. The \emph{statistical distance} between $\bD_1$ and $\bD_2$ is
$$\SD{\bD_1;\bD_2}:=\max_{T\subseteq \Omega}\left(\pr{\bD_1\in T}-\pr{\bD_2\in T}\right)=\frac{1}{2}\sum_{s\in \Omega}\abs{\bD_1(s)-\bD_2(s)}.$$ We say $\bD_1$ is $\eps$-close to $\bD_2$ if $\Delta(\bD_1;\bD_2)\le \eps$, which is also denoted by $\bD_1\approx_{\eps}\bD_2$. Specifically, when there are two joint distributions $(\bX,\bZ)$ and $(\bY,\bZ)$ such that $(\bX,\bZ)\approx_\eps(\bY,\bZ)$, we sometimes write $(\bX\approx_\eps \bY)\mid \bZ$ for short.
\end{definition}
We frequently use the following standard properties.
\begin{lemma}
For every distribution $\bD_1,\bD_2,\bD_3$ on the same universe, the following properties hold:
\begin{itemize}
\item
For any distribution $\bZ$, $\SD{(\bD_1,\bZ);(\bD_2,\bZ)}=\ex[z\sim \bZ]{\SD{\bD_1|_{\bZ=z};\bD_2|_{\bZ=z}}}$.
\item
For every function $f$, $\SD{f(\bD_1);f(\bD_2)}\le \SD{\bD_1;\bD_2}$.
\item 
$\SD{\bD_1;\bD_3}\le \SD{\bD_1;\bD_2}+\SD{\bD_2;\bD_3}$. (triangle inequality)
\end{itemize}
\end{lemma}

\subsection{Conditional Min-entropy}
\begin{definition}[\cite{DORS08}]
For a joint distribution $(\bX,\bZ)$, the \emph{average conditional min-entropy} of $\bX$ given $\bZ$ is 
$$\avghmin(\bX\mid \bZ):=-\log\left(\ex[z\sim\bZ]{\max_{x}(\pr{\bX=x\mid \bZ=z})}\right).$$
\end{definition}
The following lemma, usually 
referred to as the \emph{chain rule}, is frequently used in this paper.
\begin{lemma}[\cite{DORS08}]\label{lemma:chain-rule}
Let $\bX,\bY,\bZ$ be (correlated) random variables. Then
$$\avghmin(\bX\mid (\bY,\bZ))\ge \avghmin(\bX\mid\bZ)-\log(\Supp(\bY)).$$
\end{lemma}
When we need to consider worst-case conditional min-entropy, we use the following lemma.
\begin{lemma}[\cite{DORS08}]\label{lemma:conditional-fixed}
Let $\bX,\bZ$ be (correlated) random variables. For every $\eps>0$,
$$\pr[z\sim\bZ]{\hmin(\bX|_{\bZ=z})\ge \hmin(\bX\mid\bZ)-\log(1/\eps)}\ge 1-\eps.$$
\end{lemma}
Note that the above two lemmas imply the following:
\begin{lemma}[\cite{MW97}]\label{lemma:chain-rule-fixed}
Let $\bX,\bZ$ be (correlated) random variables. For every $\eps>0$,
$$\pr[z\sim\bZ]{\hmin(\bX|_{\bZ=z})\ge \hmin(\bX)-\log(\Supp(\bZ))-\log(1/\eps)}\ge 1-\eps.$$
\end{lemma}
\begin{lemma}[\cite{DORS08}]\label{lemma:avg-ext}
Let $\eps,\delta>0$ and $\bX,\bZ$ be a random variables such that $\avghmin(\bX\mid \bZ)\ge k+\log(1/\delta)$. Let $\Ext:\bits{n}\times\bits{d}\to\bits{m}$ be a $(k,\eps)$-seeded extractor. Then
$$\left(\Ext(\bX,\bU_d)\approx_{\eps+\delta} \bU_m\right)\mid\bZ.$$
\end{lemma}

\subsection{Branching Programs}\label{sec:branching-program}
The following definition is equivalent to \Cref{def:small-space-source} in the sense that each layer corresponds to a time step and each vertex in a layer corresponds to a state in a certain time step.
\begin{definition}
A branching program $B$ of width $w$ and length $n$ (for sampling) is a directed (multi)-graph with $(n+1)$ layers $L_0,L_1,\dots,L_n$ and has at most $w$ vertices in each layer. The first layer (indexed by $0$) has only one vertex called the start vertex, and every vertex in $L_n$ has no outgoing edge. For every vertex $v$ in layer $i<n$, the set of outgoing edges from $v$, denoted by $E_v$, satisfies the following.
\begin{itemize}
\item 
Every edge $e\in E_v$ is connected to a vertex in $L_{i+1}$. 
\item
Each edge $e\in E_v$ is labeled with a probability, denoted by $\pr{e}$, so that $\sum_{e\in E_v} \pr{e}=1$.
\item
Each edge $e\in E_v$ is labeled with a bit $b_e\in\bits{}$, and if two distinct edges $e_1,e_2\in E_v$ are connected to the same vertex $w\in L_{i+1}$ then $b_{e_1}\neq b_{e_2}$. (Note that this implies $\abs{E_v}\le 2w$.)
\end{itemize}
The output of $B$ is a $n$-bit string generated by the following process. Let $v_0$ be the start vertex. Repeat the following for $i$ from $1$ to $n$: sample an edge $e_i\in E_{v_{i-1}}$ with probability $\pr{e_i}$, output $b_{e_i}$ and let $v_i$ be the vertex which is connected by $e_i$. We say $(v_0,e_1,v_1,\dots,e_n,v_n)$ is the \emph{computation path} of $B$. We say a random variable $\bX\in\bits{n}$ is a space-$s$ source if it is generated by a branching program of width $2^s$ and length $n$. 
\end{definition}
We also consider the subprograms of a branching program. 
\begin{definition}
Let $B=(L_0,L_1,\dots,L_n)$ be a branching program of width $w$ and length $n$ and let $v$ be a vertex in layer $i$ of $B$. Then the subprogram of $B$ starting at $v$, denoted by $B_v$, is the induced subgraph of $B$ which consists of $(\{v\},L_{i+1},\dots,L_n)$. Note that $B_v$ is a branching program of width $w$ and length $n-i$ which takes $v$ as the start vertex.
\end{definition}
We need the following simple fact from \cite{KRVZ11}.
\begin{lemma}[\cite{KRVZ11}]\label{lemma:small-space-condition}
Let $\bX$ be a space-$s$ source sampled by a branching program $B$, and let $v$ be a vertex in layer $i$ of $B$. Then conditioned on the event that the computation path of $\bX$ passes $v$, $\bX$ is the concatenation of two independent random variables $\bX_1\in\bits{i}$, $\bX_2\in\bits{n-i}$. Moreover $\bX_2$ is exactly the source generated by the subprogram $B_v$.
\end{lemma}

\subsection{Seeded Extractors}\label{sec:seeded-extractors}
\begin{definition}
$\Ext:\bits{n}\times\bits{d}\to\bits{m}$ is a \emph{seeded extractor for entropy $k$ with error $\eps$} (or \emph{$(k,\eps)$-seeded extractor} for short) if for every $(n,k)$ source $\bX$, and every $\bY=\bU_d$,
$$\Ext(\bX,\bY)\approx_{\eps}\bU_m.$$
We call $d$ the \emph{seed length} of $\Ext$. We say $\Ext$ is linear if $\Ext(\cdot,y)$ is a linear function for every $y\in\bits{d}$. We say $\Ext$ is strong if 
$$(\Ext(\bX,\bY)\approx_{\eps}\bU_m)\mid \bY.$$
\end{definition}

\begin{lemma}[\cite{GUV09}]\label{lemma:GUV-ext}
There exists a constant $c_{\ref{lemma:GUV-ext}}$ and a constant $\beta>0$ such that for every $\eps>2^{-\beta n}$ and every $k$, there exists an explicit $(k,\eps)$-strong seeded extractor $\Ext:\bits{n}\times\bits{d}\to\bits{m}$ s.t. $d=c_{\ref{lemma:GUV-ext}}\log(n/\eps)$ and $m=k/2$.
\end{lemma}
We also need the following extractor from \cite{CGL21} which is linear but has worse parameters. 
\begin{lemma}\label{lemma:linear-ext}
There exists a constant $c_{\ref{lemma:linear-ext}}$ such that for every $t,m\in\bbN$ and $\eps>0$, there exists an explicit $(c_{\ref{lemma:linear-ext}}(m+\log(1/\eps)),\eps)$-linear strong seeded extractor $\LExt:\bits{n}\times\bits{d}\to\bits{m}$ s.t. $d=O(\frac{m}{t}+\log(n/\eps)+\log^2(t)\log(m/\eps))$.
\end{lemma}
Note that when $m=t\log(n/\eps)$ the seed length is bounded by $O\left(\left(\log^2(t)+1\right)\log(n/\eps))\right)$. 

\subsection{Samplers}
First we define a sampler. We note that the our definition is different from the standard definition of averaging samplers~\cite{BR94} in the following sense: first, we need the sampler to work even when the given randomness is only a weak source. Second, we only care about ``small tests".
\begin{definition}
$\Samp:\bits{n}\times[D]\to\bits{m}$ is a \emph{$(\eps,\delta)$-sampler} for entropy $k$ if for every set $T\subseteq \bits{m}$ s.t. $|T|\le \eps 2^m$ and every $(n,k)$-source $\bX$,
$$\pr[x\sim\bX]{\pr[y\sim{[D]}]{\Samp(x,y)\in T}>2\eps}\le \delta.$$
We say $\Samp$ is linear if $\Samp(\cdot,y)$ is linear for every $y\in[D]$. 
\end{definition}
Zuckerman~\cite{Zuc97} showed that one can use a seeded extractor as a sampler for weak sources.
\begin{lemma}[\cite{Zuc97}]\label{lemma:Zuc97}
A $(k+\log(1/\delta),\eps)$-seeded extractor is also a $(\eps,\delta)$-sampler for entropy $k$.
\end{lemma}
The following is a relaxation of a sampler, which is called a somewhere random sampler.
\begin{definition}
$\Samp:\bits{n}\times[D]\times[C]\to\bits{m}$ is a \emph{$(\eps,\delta)$-somewhere random sampler} for entropy $k$ if for every set $T\subseteq \bits{m}$ s.t. $\abs{T}\le \eps 2^m$ and every $(n,k)$-source $\bX$,
$$\pr[x\sim\bX]{\pr[y\sim{[D]}]{\forall z\in[C] \,\,\Samp(x,y,z)\in T}>2\eps}\le \delta.$$
We say $\Samp$ is linear if $\Samp(\cdot,y,z)$ is linear for every $y\in[D],z\in[C]$. 
\end{definition}
The following lemma is implicit in \cite{BDT17}. For completeness we include a proof in \Cref{appendix:BDT-proof}.
\begin{lemma}[\cite{BDT17}]\label{lemma:BDT}
If there exists an explicit $(\eps,\delta)$-sampler $\Samp:\bits{n}\times[D_0]\to\bits{m}$ for entropy $k$, then for every constant $\gamma<1$ there exists an explicit $(D^{-\gamma},\delta)$-somewhere random sampler $\Samp':\bits{n}\times[D]\times[C]\to\bits{m}$ for entropy $k$ with $D=D_0^{O(1)}$ and $C=O\left(\frac{\log(D_0)}{\log(1/\eps)}\right)$. Furthermore if $\Samp$ is linear then $\Samp'$ is also linear. 
\end{lemma}

By \Cref{lemma:linear-ext}, \Cref{lemma:Zuc97} and \Cref{lemma:BDT} we can get the following explicit somewhere random smapler.
\begin{lemma}\label{lemma:sr-sampler}
For every constant $\gamma<1$, and every $\delta>0,t<2^{\sqrt[3]{\log(n)}}$ there exists an explicit $(D^{-\gamma},\delta)$-linear somewhere random sampler $\Samp:\bits{n}\times[D]\times[C]\to\bits{t\log(n)}$ for entropy $O(t\log(n))+\log(1/\delta)$, where $D=n^{O(1)}$ and $C=O(\log^2(t))$.
\end{lemma}
\begin{proof}
By \Cref{lemma:linear-ext} and \Cref{lemma:Zuc97}, there exists an explicit $(\eps,\delta)$-linear sampler $\Samp':\bits{n}\times [D_0]\to\bits{t\log(n)}$ for entropy $O(t\log(n))+\log(1/\delta)$ where $\eps=2^{-\log(n)/\log^2(t)}$ and $D_0=n^{O(1)}$. The claim follows by applying \Cref{lemma:BDT} on $\Samp'$.
\end{proof}

\subsection{Non-Oblivious Bit-Fixing Sources}
\begin{definition}\label{def:nobf}
A distribution $\bX=(\bX_1,\bX_2,\dots,\bX_n)$ on $\bits{n}$ is called $t$-wise independent if for every subset $S\subseteq [n]$ of size $t$ we have
$\bX_S=\bU_q.$
\end{definition}

\begin{lemma}[\cite{AGM03}]\label{lemma:AGM}
Let $\bX=(\bX_1,\bX_2,\dots,\bX_n)$ be a distribution on $\bits{n}$. If for every $S\subseteq [n]$ s.t. $\abs{S}\le t$,
$$\bigoplus_{i\in S} \bX_i \approx_{\gamma} \bU_1,$$
then $\bX$ is $2n^{t}\gamma$-close to a $t$-wise independent distribution.
\end{lemma}

\begin{definition}
A distribution $\bX=(\bX_1,\bX_2,\dots,\bX_n)$ on $\bits{n}$ is called a $(q,t)$-non-oblivious bit-fixing (NOBF) source if there exists a set $Q$ s.t. $\abs{Q}\le q$ and $\bX_{[n]\backslash Q}$ is $t$-wise independent.
\end{definition}

In this paper we need the following extractors for NOBF sources. 
\begin{lemma}[\cite{CZ19,Li16}]\label{lemma:NOBF-ext-poly-error}
There exists an explicit function $\BFExt:\bits{n}\to\bits{m}$ for $(q,t)$-NOBF sources with error $n^{-\Omega(1)}$ where $m=n^{\Omega(1)}$, $q=n^{0.9}$ and $t=(m\log(n))^{C_{\ref{lemma:NOBF-ext-poly-error}}}$ for some constant $C_{\ref{lemma:NOBF-ext-poly-error}}$.
\end{lemma}
\begin{lemma}[\cite{Vio14}]\label{lemma:NOBF-ext-const-error}
For every $\eps>0$, the majority function $\Maj:\bits{n}\to\bits{}$ is an extractor for $(q,t)$-NOBF sources with error $\eps+O(n^{-0.1})$ where $q=n^{0.4}$ and $t=O(\eps^{-2}\log^2(1/\eps))$.
\end{lemma}

\subsection{Markov Chain}
In this paper we usually consider the scenario that we have two sources $\bX,\bY$ which are independent conditioned on a collection of random variables $\bZ$. We use Markov chain as a shorthand for this.
\begin{definition}
Let $\bX,\bY,\bZ$ be random variables. We say $\bX \lra \bZ \lra \bY$ is a Markov chain if $\bX$ and $\bY$ are independent conditioned on any fixing of $\bZ$.
\end{definition}
We frequently use the following fact. 
\begin{lemma}\label{lemma:two-party}
If $\bX\lra \bZ\lra \bY$ is a Markov chain, then for every deterministic function $f$, let $\bW=f(\bX,\bZ)$. Then
\begin{itemize}
\item
$(\bX,\bW)\lra \bZ\lra \bY$ is a Markov chain.
\item
$\bX\lra (\bW,\bZ)\lra \bY$ is a Markov chain.
\end{itemize}
We use ``$\bW$ is a deterministic function of $\bX$ (conditioned on $\bZ$)" to refer to the first item, and ``fix $\bW$" to refer to the second item.
\end{lemma}
\subsection{Independence Merging}\label{sec:ind-merging}
The following lemma is from \cite{CGL21} and is based on the ideas in \cite{CL16-NIPM}. Basically it says that if $\bY$ is independent of some tampered seeds $\bY^S$, and $\bX$ has enough entropy when conditioned some tampered sources $\bX^T$, then a strong seeded extractor can ``merge" the independence of $\bY$ from $\bY^S$ and $\bX$ from $\bX^T$.
\begin{lemma}[independence-merging lemma]\label{lemma:ind-merging}
Let $(\bX,\bX^{[t]})\lra \bZ\lra (\bY,\bY^{[t]})$ be a Markov chain, such that $\bX,\bX^{[t]}\in\bits{n}$, $\bY,\bY^{[t]}\in\bits{d}$. Moreover, suppose there exists $S,T\subseteq [t]$ such that
\begin{itemize}
\item
$(\bY\approx_{\delta}\bU_d)\mid(\bZ,\bY^S)$
\item
$\avghmin(\bX\mid (\bX^T,\bZ))\ge k+tm+\log(1/\eps)$
\end{itemize}
Let $\Ext:\bits{n}\times \bits{d}\to\bits{m}$ be any $(k,\eps)$-strong seeded extractor, let 
$\bW=\Ext(\bX,\bY)$ and 
$\bW^j=\Ext(\bX^j,\bY^j)$ for every $j\in[t]$.
Then
$$(\bW\approx_{2\eps+\delta} \bU_m)\mid(\bW^{S\cup T},\bY,\bY^{[t]},\bZ).$$
\end{lemma}

\subsection{Basic Properties in Additive Combinatorics}\label{sec:additive-combinatorics}
\begin{definition}
For every two sets $A,B\subseteq \bbF_2^n$, we define $A+B=\{a+b:a\in A, b\in B\}$. For $b\in\bbF_2^n$ we use $A+b$ as the shorthand for $A+\{b\}$. For every $\ell \in \bbN$ and every $A\subseteq \bbF_2^n$, define $1A=A$ and $\ell A = A+(\ell-1)A$ recursively.
\end{definition}

\begin{lemma}[\cite{Plunnecke,Ruzsa99}]\label{lemma:Plunnecke}
For every $A,B\subseteq \bbF_2^n$ s.t. $\abs{A}=\abs{B}$ and $\abs{A+B}\le r \abs{A}$, $\abs{kA+\ell B}\le r^{k+\ell+1}\abs{A}$ for every $k,\ell\in\bbN$.
\end{lemma}
\begin{definition}
We say a function $\phi:\bbF_2^n\to\bbF_2^m$ is a $s$-Freiman homomorphism of a set $A\subseteq \bbF_2^n$ if for every $a_1,\dots,a_s,a_1',\dots,a_s'\in A$, 
$$\phi(a_1)+\ldots+\phi(a_s)=\phi(a_1')+\ldots+\phi(a_s')\Rightarrow a_1+\ldots+a_s = a_1'+\ldots+a_s'.$$
\end{definition}
The following property is easy to verify. 
\begin{lemma}\label{lemma:linear-Freiman}
If $\phi$ is a linear $s$-Freiman homomorphism, then $\phi$ is injective on $sA+v$ for every $v\in\bbF_2^n$. Further, for $x\in 2sA$ we have $\phi(x)=0\Leftrightarrow x=0$.
\end{lemma}
The following lemma can be used to obtain a linear Freiman homomorphism with small image. 
\begin{lemma}[\cite{GR07}]\label{lemma:Green-Rusza}
For every set $A\subseteq \bbF_2^n$ there exists a linear $s$-Freiman homomorphism $\phi:\bbF_2^n\to\bbF_2^m$ of $A$ such that $\phi(2sA)=\bbF_2^m$.
\end{lemma}

\subsection{Fourier Analysis}
First we recall some basic definitions and properties in Fourier analysis.
\begin{definition}
Let $f:\bbF_2^n\to \bbR$ be a function. The \emph{Fourier coefficients} of $f$, denoted by $\widehat{f}:\bbF_2^n\to\bbR$, are $$\widehat{f}(\alpha):=\ex[x\sim \bbF_2^n]{f(x)\cdot (-1)^{\inp{\alpha,x}}}.$$
\end{definition}
\begin{lemma}[Parseval-Plancherel identity]\label{lemma:Parseval}
For every functions $f,g:\bbF_2^n\to\bbR$, 
$$\ex[x\sim \bbF_2^n]{f(x)g(x)}=\sum_{\alpha\in\bbF_2^n} \widehat{f}(\alpha)\widehat{g}(\alpha).$$
\end{lemma}
\begin{definition}
The convolution of functions $f,g:\bbF_2^n\to \bbR$, denoted by $f*g:\bbF_2^n\to\bbR$, is defined as 
$$f*g(x):=\ex[y\sim\bbF_2^n]{f(y)g(x-y)}.$$
\end{definition}
\begin{lemma}
For every functions $f,g:\bbF_2^n\to\bbR$ and every $\alpha\in\bbF_2^n$, 
$$\widehat{f*g}(\alpha)=\widehat{f}(\alpha)\widehat{g}(\alpha).$$
\end{lemma}
Next we define a density function.
\begin{definition}
For every $A\subseteq \bbF_2^n$, define the density function of $A$ to be $\mu_A:=\frac{2^n}{\abs{A}}\cdot \mathds{1}_A$. For a distribution $\bA$ on $\bbF_2^n$, the density function of $\bA$, denoted by $\mu_{\bA}$, is defined as $\mu_{\bA}(x)=2^n\pr{\bA=x}$.
\end{definition}
We need the following three facts about density functions.
\begin{lemma}\label{lemma:expectation}
Let $f:\bbF_2^n\to\bbR$ be a function and let $\bA$ be a distribution on $\bbF_2^n$. Then $$\ex{f(\bA)}=\ex[x\sim \bbF_2^n]{\mu_{\bA}(x) f (x)}.$$
\end{lemma}
\begin{lemma}\label{lemma:convolution}
Let $\bA,\bB$ be two distributions on $\bbF_2^n$. Then $\mu_{\bA+\bB}=\mu_{\bA}*\mu_{\bB}$.
\end{lemma}
\begin{lemma}\label{lemma:density-subspace}
If $V\subseteq \bbF_2^n$ is a linear subspace, then $\widehat{\mu_V}(\alpha)=1$ if $\alpha\in V^{\bot}$ and $\widehat{\mu_V}(\alpha)=0$ otherwise.
\end{lemma}
Finally we need Chang's lemma.
\begin{lemma}[\cite{Chang}]\label{lemma:Chang}
For $X\subseteq \bbF_2^n$, define $\Spec_{\gamma}(X)=\{\alpha\in\bbF_2^n:\abs{\widehat{\mu_X}(\alpha)}\ge \gamma\}$. Define $\beta=\abs{X}/\abs{\bbF_2^n}$. Then 
$$\dim(\linspan(\Spec_{\gamma}(X)))\le 2\gamma^{-2}\ln(1/\beta).$$
\end{lemma}
\subsection{Minimax Theorem}\label{sec:minimax}
\begin{lemma}[minimax theorem~\cite{vN28}]
Let $\cX\subseteq \bbR^n,\cY\subseteq \bbR^m$ be convex sets. Then for every bilinear function $g:\cX\times\cY\to\bbR$, 
$$\min_{x\in\cX}\max_{y\in\cY} g(x,y)=\max_{y\in \cY}\min_{x\in \cX} g(x,y).$$
\end{lemma}
\begin{corollary}\label{coro:distribution-minimax}
Let $\Omega$ be a finite set, $\cX$ be a convex set of distributions on $\Omega$, and $\bY$ be a distribution on $\Omega$. If for every function $f:\Omega\to{[0,1]}$ there exists $\bX_f\in\cX$ such that $\ex{f(\bX_f)}-\ex{f(\bY)}\le \eps$, then there exists $\bX^*\in \cX$ such that $\bY\approx_\eps \bX^*$.
\end{corollary}
\begin{proof}
Let $\cF$ denote the set of all the functions from $\Omega$ to $[0,1]$. Note that a distribution $\bX$ can be represented by a vector in $\bbR^{\abs{\Omega}}$, where the coordinate indexed by $s\in\Omega$ is $\pr{\bX=s}$. A function $f:\Omega\to{[0,1]}$ can also be represented by a vector in $\bbR^{\abs{\Omega}}$, where the coordinate indexed by $s\in\Omega$ is $f(s)$. Observe that $\cF$ is convex. Define the function $g:\cX\times \cF\to \bbR$ to be 
$$g(\bX,f):=\ex{f(\bX)}-\ex{f(\bY)}=\left(\sum_{s\in \Omega}\pr{\bX=s}\cdot f(s)\right) - \ex{f(\bY)}.$$
Observe that $g$ is bilinear. By minimax theorem, 
$$\min_{\bX\in\cX}\max_{f\in \cF} g(\bX,f) = \max_{f\in\cF} \min_{\bX\in\cX} g(\bX,f) \le \max_{f\in\cF} (\ex{f(\bX_f)}-\ex{f(\bY)})\le \eps.$$
That is, there exists $\bX^*\in\cX$ such that for every function $f:\Omega\to{[0,1]}$, $\ex{f(\bX^*)}-\ex{f(\bY)}\le \eps$. If we take $f=\mathds{1}_T$ for some $T\subseteq \Omega$, then $\ex{f(\bX^*)}-\ex{f(\bY)}$ is exactly $\pr{\bX^*\in T}-\pr{\bY\in T}$. Therefore by definition of statistical distance, $\bX^*\approx_{\eps} \bY$. 
\end{proof}
\section{Improved Reduction for Small-Space Sources}\label{sec:small-space}
In this section we prove the following lemma.
\begin{lemma}\label{lemma:small-space-reduction}
For every integer $C\ge 2$, every space-$s$ source on $n$-bit with min-entropy $$k'\ge Ck+(C-1)\left(2s+2\log(n/\eps)\right)$$ is $(3C\eps)$-close to a convex combination of $(n,k,C)$-sumset sources. 
\end{lemma}
Note that by taking $C=2$ in \Cref{lemma:small-space-reduction}, we can prove that the sumset source extractor in \Cref{thm:ext-poly-error} and \Cref{thm:ext-const-error} are also small-space source extractors which satisfy the parameters in \Cref{thm:small-space-ext-poly-error} and \Cref{thm:small-space-ext-const-error} respectively. In the rest of this section we focus on proving \Cref{lemma:small-space-reduction}. First we show how to prove \Cref{lemma:small-space-reduction} based on the following lemma.
\begin{lemma}\label{lemma:small-space-two-part}
Every space-$s$ source $\bX\in\bits{n}$ with entropy at least $k=k_1+k_2+2s+2\log(n/\eps)$ is $3\eps$-close to a convex combination of sources of the form $\bX_1\circ\bX_2$ which satisfy the following properties:
\begin{itemize}
\item 
$\bX_1$ is independent of $\bX_2$
\item
$\hmin(\bX_1)\ge k_1$, $\hmin(\bX_2)\ge k_2$
\item
$\bX_2$ is a space-$s$ source
\end{itemize}
\end{lemma}
\begin{proof}[Proof of \Cref{lemma:small-space-reduction}]
By induction, \Cref{lemma:small-space-two-part} implies that a space-$s$ source with entropy $Ck+(C-1)(2s+2\log(n/\eps))$ is $3C\eps$-close to a convex combination of sources of the form $\bX_1\circ\bX_2\circ\dots\circ\bX_C$ such that $\bX_1,\ldots,\bX_C$ are independent, and for every $i\in[C]$, $\hmin(\bX_i)\ge k$. Let $\ell_1,\ell_2,\dots,\ell_C$ denote the length of $\bX_1,\bX_2,\dots,\bX_C$ respectively and define $p_i=\sum_{j=1}^{i-1} \ell_i$ and $s_i=\sum_{j=i+1}^n \ell_j$ (note that $p_1=0$ and $s_C=0$). Then observe that 
$$\bX_1\circ\dots\circ \bX_C=\sum_{i=1}^C 0^{p_i}\circ \bX_i \circ 0^{s_i},$$
which implies that $\bX=\bX_1\circ\dots\circ \bX_C$ is a $(n,k,C)$-sumset source. 
\end{proof}
To prove \Cref{lemma:small-space-two-part}, first we need the following lemma. 
\begin{lemma}\label{lemma:entropy-drop}
Let $B$ be a branching program of width $2^s$ and length $n$ for sampling. Let $e$ be an edge in $B$ connected from $u$ to $v$ and let $\bX_u,\bX_v$ be the output distributions of the subprograms $B_u,B_v$ respectively. Then $\hmin(\bX_v)\ge \hmin(\bX_u)-\log(1/\pr{e})$.
\end{lemma}
\begin{proof}
Let $x^*=\arg\max_{x}\pr{\bX_v=x}$. Note that $\hmin(\bX_v)=-\log(\pr{\bX_v=x^*})$ by definition. Observe that $\pr{\bX_u=b_e\circ x^*}\ge \pr{e}\cdot \pr{\bX_v=x^*}$. Therefore, $$\hmin(\bX_u)\le -\log\left(\pr{\bX_u=b_e\circ x^*}\right)\le -\log\left(\pr{e}\cdot \pr{\bX_v=x^*}\right)\le \hmin(\bX_v)+\log(1/\pr{e}).$$
\end{proof}
Next we prove \Cref{lemma:small-space-two-part}.
\begin{proof}[Proof of \Cref{lemma:small-space-two-part}]
Let $B$ denote the branching program which samples $\bX$. For every $v$, define $\bX_v$ to be the source generated by the subprogram $B_v$. Define $v$ to be a \emph{stopping vertex} if $\hmin(\bX_v)\le k_2+s+\log(n/\eps)$. Observe that every vertex $u$ in the last layer is a stopping vertex since $\hmin(\bX_u)=0$, so there is always a stopping vertex in the computation path. We define an edge $e$ in $B$ to be a \emph{bad edge} if $\pr{e}\le\eps/(n\cdot 2^s)$. Now define a random variable $\bV$ as follows:
\begin{itemize}
\item 
$\bV=\bot$ if the computation path of $\bX$ visits a bad edge before visiting any stopping vertex,
\item
otherwise, $\bV=v$ where $v$ is the first stopping vertex in the computation path.
\end{itemize}
Observe that $\pr{\bV=\bot}\le 2\eps$, since in each step of $B$ there are at most $2^{s+1}$ edges starting from the current vertex, and there are $n$ steps in total. Define 
$$\mathsf{BAD}=\{v\in \Supp(\bV): \hmin(\bX|_{\bV=v})\le k-s-\log(n/\eps)\}.$$
Then $\pr{\bV\in\mathsf{BAD}}\le \eps$ by \Cref{lemma:chain-rule-fixed}. We claim that if $v\not\in\mathsf{BAD}$ and $v\neq \bot$, then conditioned on $\bV=v$, the source $\bX$ can be written as $\bX_1\circ \bX_2$ which satisfies the properties stated in \Cref{lemma:small-space-two-part}. The claim directly implies \Cref{lemma:small-space-two-part} because $\pr{v\in\mathsf{BAD}\vee v=\bot}\le 3\eps$ by union bound. Next we prove the claim. Let $E_1$ denote the event ``the computation path contains $v$", and $E_2$ denote the event ``the computation path does not contain any bad edge or stopping vertex before the layer of $v$". Observe that $\bV=v$ is equivalent to $E_1\wedge E_2$. Conditioned on $E_1$, by \Cref{lemma:small-space-condition}$, \bX$ can be written as $\bX_1\circ\bX_2$ where $\bX_1$ is independent of $\bX_2$ and $\bX_2=\bX_v$. Now observe that $E_2$ only involves layers before $v$, so conditioned on $E_1$, $\bX_2$ is independent of $E_2$. Therefore, conditioned on $\bV=v$, we still have $\bX_2=\bX_v$, which is a space-$s$ source, and $\bX_1$ is still independent of $\bX_2$. Next observe that $$\hmin(\bX_1)=\hmin(\bX|_{\bV=v})-\hmin(\bX_2)\ge (k-s-\log(n/\eps))-(k_2+s+\log(n/\eps))\ge k_1.$$ It remains to prove that $\hmin(\bX_2)\ge k_2$. Assume for contradiction that $\hmin(\bX_v)<k_2$. Let $e$ be the edge in the computation path which connects to $v$, and suppose $e$ is from $u$. Now consider the following two cases.
\begin{itemize}
\item
If $e$ is not a bad edge, then $\hmin(\bX_u)\le \hmin(\bX_v)+\log(1/\pr{e})<k_2+s+\log(n/\eps)$, which means $u$ is also a stopping vertex. Therefore $v$ cannot be the first stopping vertex.
\item
If $e$ is a bad edge, then $\bV=\bot$. 
\end{itemize}
In both cases $\bV\neq v$, which is a contradiction. In conclusion we must have $\hmin(\bX_2)\ge k_2$. 
\end{proof}
\section{Extractors for Sum of Two Sources}\label{sec:sumset-ext}
In this section we formally prove \Cref{thm:ext-poly-error} and \Cref{thm:ext-const-error}. The construction of our extractors relies on the following lemma:
\begin{lemma}[main lemma]\label{lemma:main}
For every constant $\gamma<1$ and every $t\in\bbN$, there exists $N=n^{O(1)}$ and an explicit function $\Reduce:\bits{n}\to\bits{N}$ s.t. for every $(n,k,2)$-sumset source $\bX$, where $$k=O\left(t^3\log\left(n\right)\cdot\left(\frac{\log\log(n)}{\log\log\log(n)}+\log^3(t)\right)\cdot\left(\log\log\log^4(n)+\log^4(t)\right)\right),$$ $\Reduce(\bX)$ is $N^{-\gamma}$-close to a $(N^{1-\gamma},t)$-NOBF source.
\end{lemma}
Before we prove \Cref{lemma:main}, first we show how to prove \Cref{thm:ext-poly-error} and \Cref{thm:ext-const-error}  based on \Cref{lemma:main}. 
\begin{proof}[Proof of \Cref{thm:ext-poly-error}]
Let $\Reduce:\bits{n}\to\bits{N}$ be the function from \Cref{lemma:main} by taking $\gamma=0.1$. Note that $N=\poly(n)$. Let $\BFExt:\bits{N}\to\bits{m}$ be the NOBF-source extractor from \Cref{lemma:NOBF-ext-poly-error}. Let $\bX$ be a $(n,k,2)$-source, where $k$ is defined later. If $\Reduce(\bX)$ is $N^{-\Omega(1)}$-close to a $(N^{0.9},t)$-NOBF source where $t=(m\log(N))^{C_{\ref{lemma:NOBF-ext-poly-error}}}$, then $$\Ext(\bX):=\BFExt(\Reduce(\bX))$$
is $n^{-\Omega(1)}$-close to uniform. By \Cref{lemma:main} it suffices to take $k=O(t^3\log^7(t)\log(n))\le(m\log(n))^{1+3C_{\ref{lemma:NOBF-ext-poly-error}}}.$
\end{proof}
\begin{proof}[Proof of \Cref{thm:ext-const-error}]
Let $\Reduce:\bits{n}\to\bits{N}$ be the function from \Cref{lemma:main} by taking $\gamma=0.6$. Note that $N=\poly(n)$. Let $\Maj:\bits{N}\to\bits{}$ be the NOBF-source extractor from \Cref{lemma:NOBF-ext-const-error}, i.e. the majority function. Let $\bX$ be a $(n,k,2)$-source, where $k$ is defined later. If $\Reduce(\bX)$ is $(\eps/2)$-close to a $(N^{0.4},t)$-NOBF source where $t=O(\eps^{-2}\log^2(1/\eps))=O(1)$, then $$\Ext(\bX):=\Maj(\Reduce(\bX))$$
is $\eps$-close to uniform. By \Cref{lemma:main} it suffices to take $k=O(\log(n)\log\log(n)\log\log\log^3(n)).$
\end{proof}

Next we prove \Cref{lemma:main}. First we recall the definition of a strong affine correlation breaker. To simplify our proof of \Cref{lemma:main}, here we use a definition which is slightly more general than \Cref{def:strong-ACB-informal}.
\begin{definition}\label{def:strong ACB}
$\ACB:\bits{n}\times\bits{d}\times\bits{a}\to\bits{m}$ is a $(t,k,\gamma)$-affine correlation breaker if for every distribution $\bX,\bA,\bB\in\bits{n}$, $\bY,\bY^{[t]}\in\bits{d}$, $\bZ$ and string $\alpha,\alpha^{[t]}\in\bits{a}$ s.t.
\begin{itemize}
\item
$\bX=\bA+\bB$
\item 
$\avghmin(\bA\mid\bZ)\ge k$
\item
$(\bY,\bZ)=(\bU_d,\bZ)$
\item
$\bA\lra\bZ\lra (\bB,\bY,\bY^{[t]})$ is a Markov chain
\item
$\forall i\in[t]$, $\alpha\neq \alpha^{i}$
\end{itemize}
It holds that 
$$(\ACB(\bX,\bY,\alpha)\approx_{\gamma}\bU_m)\mid \left(\{\ACB(\bX,\bY^{i},\alpha^{i})\}_{i\in[t]},\bZ\right).$$
We say $\ACB$ is strong if 
$$(\ACB(\bX,\bY,\alpha)\approx_{\gamma}\bU_m)\mid \left(\{\ACB(\bX,\bY^{i},\alpha^{i})\}_{i\in[t]},\bY,\bY^{[t]},\bZ\right).$$
\end{definition}

To prove \Cref{lemma:main}, we need the following lemma, which is an analog of \cite[Lemma 2.17]{CZ19}. Roughly speaking, we show that even if the seeds of the correlation breaker are added by some leakage from the source, most of the seeds are still good. 
\begin{lemma}\label{lemma:strongness-with-shifting}
For every error parameter $\gamma>0$ the following holds. Let
\begin{itemize}
\item 
  $\ACB:\bits{n}\times\bits{d}\times\bits{a}\to\bits{m}$ be a $(t,k,\eps)$-strong affine correlation breaker
  \item
  $L:\bits{n}\times\bits{a}\to\bits{d}$ be any deterministic function, which we call the \emph{leakage function}
  \item
  $\alpha,\alpha^{[t]}$ be any $a$-bit \emph{advice} s.t. $\alpha\neq \alpha^i$ for every $i\in[t]$
  \item
  $\bA$ be a $(n,k+(t+1)\ell)$-source
\end{itemize} For every $b\in\bits{n}$, $y\in\bits{d}$, define $$\bR_{b,y}:=\ACB(\bA+b,y+L(\bA,\alpha),\alpha)$$ and for every $i\in[t]$ define $$\bR_{b,y}^i:=\ACB(\bA+b,y+L(\bA,\alpha^{i}),\alpha^i).$$  Define
$$\mathsf{BAD}_{\alpha,\alpha^{[t]}}:=\left\{y\in\bits{d}:\exists b,y^{[t]}\textrm{ s.t. }(\bR_{b,y}\not\approx_{\gamma}\bU_m)\mid \{\bR_{b,y^i}^i\}_{i\in[t]}\right\},$$
which denotes the ``bad seeds" of $\ACB$ determined by $\bA$, $L$ and $\alpha,\alpha^{[t]}$. Then
$$\pr[y\sim \bU_d]{y\in\mathsf{BAD}_{\alpha,\alpha^{[t]}}}\le \frac{\eps}{\gamma}.$$
\end{lemma}
\begin{proof}
Define deterministic functions $f^1,\dots,f^t:\bits{d}\to\bits{d}$ and $g:\bits{d}\to\bits{n}$ s.t. for every $y\in\mathsf{BAD}_{\alpha,\alpha^{[t]}}$, 
$$\left(\bR_{g(y),y}\not\approx_{\gamma}\bU_m\right)\mid \left(\{\bR_{g(y),f^i(y)}^i\}_{i\in[t]}\right).$$
For $y\not\in\mathsf{BAD}_{\alpha,\alpha^{[t]}}$ the values of $f^1(y),f^2(y),\dots,f^t(y),g(y)$ are defined arbitrarily. Note that the existence of $f^{1},\dots,f^t,g$ is guaranteed by the definition of $\mathsf{BAD}_{\alpha,\alpha^{[t]}}$. Let $\bW:=\bU_d$ and $\delta:=\pr{\bW\in\mathsf{BAD}_{\alpha,\alpha^{[t]}}}$. Observe that 
$$(\bR_{g(\bW),\bW}\not\approx_{\gamma\delta}\bU_m)\mid (\{\bR_{g(\bW),f^i(\bW)}^i\}_{i\in[t]},\bW).$$ Now define $\bY:=\bW+L(\bA,\alpha)$, $\bY^i:=\bW+L(\bA,\alpha^i)$ for every $i\in[t]$ and $\bB:=g(\bW)$. Let $\bZ:=(L(\bA,\alpha),L(\bA,\alpha^1),\dots,L(\bA,\alpha^t))$. Note that $\bZ\in\bits{(t+1)\ell}$ is a deterministic function of $\bA$. With these new definitions the above equation can be rewritten as 
\begin{equation}\label{eq:error-lb}
(\ACB(\bA+\bB,\bY,\alpha)\not\approx_{\gamma\delta}\bU_m)\mid (\{\ACB(\bA+\bB,\bY^{i},\alpha^{i}\}_{i\in[t]},\bW).
\end{equation}
Next, observe that the following conditions hold:
\begin{itemize}
\item 
$\avghmin(\bA\mid \bZ)\ge k$ (by \Cref{lemma:chain-rule})
\item
$(\bY,\bZ)=(\bU_d,\bZ)$.
\item
$\bA\lra \bZ\lra (\bB,\bY,\bY^{[t]})$ is a Markov chain.
\end{itemize}
Note that the last condition holds because $\bZ$ is a deterministic function of $\bA$, which implies 
$\bA\lra \bZ\lra (\bB,\bW),$
and $\bY,\bY^{[t]}$ are deterministic functions of $(\bZ,\bW)$. By the definition of $\ACB$ we have 
$$(\ACB(\bA+\bB,\bY,\alpha)\approx_{\eps}\bU_m)\mid (\{\ACB(\bA+\bB,\bY^{i},\alpha^{i}\}_{i\in[t]},\bY,\bZ)$$ which implies 
\begin{equation}\label{eq:error-ub}
(\ACB(\bA+\bB,\bY,\alpha)\approx_{\eps}\bU_m)\mid (\{\ACB(\bA+\bB,\bY^{i},\alpha^{i}\}_{i\in[t]},\bW)
\end{equation}
since $\bW=\bY-L(\bA,\alpha)$ and $L(\bA,\alpha)$ is a part of $\bZ$.
By (\ref{eq:error-lb})  and (\ref{eq:error-ub}) we get $\delta \le \eps/\gamma$.
\end{proof}
Next we prove the following lemma, which directly implies \Cref{lemma:main} by plugging in proper choices of somewhere random samplers and affine correlation breakers.
\begin{lemma}\label{lemma:compose}
For every $\eps,\delta>0$ the following holds. Let $\ACB:\bits{n}\times\bits{d}\times[AC]\to\bits{}$ be a $(Ct-1)$-strong affine correlation breaker for entropy $k_1$ with error $A^{-2t}C^{-1}\eps\delta$, and let $\Samp:\bits{n}\times[A]\times[C]\to\bits{d}$ be a $(\eps,\delta)$-somewhere random sampler for entropy $k_2$. Then for every $n$-bit source $\bX=\bX_1+\bX_2$ such that 
$\bX_1$ is independent of $\bX_2$, 
$\hmin(\bX_1)\ge k_1+Ctd$ and $\hmin(\bX_2)\ge k_2$,
the source 
$$\Reduce(\bX):=\left\{\bigoplus_{z\in[C]}\ACB\left(\bX,\Samp(\bX,\alpha,z),(\alpha, z)\right)\right\}_{\alpha\in[A]}$$
is $3\delta$-close to a convex combination of $(2\eps A, t)$-NOBF source.
\end{lemma}
\begin{proof}
Consider \Cref{lemma:strongness-with-shifting} by taking $\bX_1$ as the source, 
$A^{-t}\delta$ as the error parameter and $L(x,(\alpha,z)):=\Samp(x,\alpha,z)$ as the leakage function. For every non-empty subset $T\subseteq [A]$ of size at most $t$ and every $z^*\in[C]$, define a set $\mathsf{BAD}'_{T,z^*}$ as follows. Let $\alpha^*$ denote the first element in $T$. Let $\beta=(\alpha^*, z^*)$ and $$\beta'=\{(\alpha,z)\}_{\alpha\in T, z\in [C]}\backslash \{\beta\}.$$ Note that $\beta'$ contains at most $2^ct-1$ advice which are all different from $\beta$. Then we define 
$$\mathsf{BAD}'_{T,z^*}:=\mathsf{BAD}_{\beta,\beta'},$$
where $\mathsf{BAD}_{\beta,\beta'}$ is defined as in \Cref{lemma:strongness-with-shifting}. Observe that by definition of $\mathsf{BAD}'_{T,z^*}$, for every $x_2\in\bits{n}$, if $\Samp(x_2,\alpha^*,z^*)\not\in \mathsf{BAD}'_{T,z^*}$, then 
$$
\left(\bigoplus_{\alpha\in T}\bigoplus_{z\in[C]}\ACB\left(\bX_1+x_2,\Samp(\bX_1,\alpha,z)+\Samp(x_2,\alpha,z),(\alpha, z)\right)\right)\approx_{A^{-t}\delta} \bU_1.
$$
By the linearity of $\Samp$, we know that for every fixing $\bX_2=x_2$, if $\Samp(x_2,\alpha^*,z^*)\not\in \mathsf{BAD}'_{T,z^*}$, then
\begin{equation}\label{eq:small-bias}
\left(\bigoplus_{\alpha\in T}\bigoplus_{z\in[C]}\ACB\left(\bX,\Samp(\bX,\alpha,z),(\alpha,z)\right)\right)\approx_{A^{-t}\delta} \bU_1.
\end{equation}

By \Cref{lemma:strongness-with-shifting} we know that $\pr[y\sim \bU_d]{y\in\mathsf{BAD}'_{T,z^*}}\le A^{-t}C^{-1}\eps$. Now define $\mathsf{BAD}'$ to be the union of $\mathsf{BAD}'_{T,z^*}$ for all possible choices of $T,z^*$. Since there are at most $A^t$ choices of $T$ and $C$ choices of $z^*$, by union bound we know that $\pr[y\sim \bU_d]{y\in\mathsf{BAD}'}\le \eps$. Therefore, by definition of somewhere random sampler, 
$$\pr[x_2\sim \bX_2]{\abs{\{\alpha\in[A]: \forall z \,\,\Samp(x_2,\alpha,z)\in \mathsf{BAD}'\}}\le 2\eps A}\ge 1-\delta.$$
In other words, with probability at least $1-\delta$ over the fixing $\bX_2=x_2$, there exists a set $Q\subseteq [A]$ of size at most $2\eps A$ which satisfies the following: for every $\alpha\in [A]\backslash Q$, there exists $z_\alpha$ such that $\Samp(x_2,\alpha,z_\alpha)\not\in \mathsf{BAD}'$, which also implies $\Samp(x_2,\alpha,z_\alpha)\not\in\mathrm{BAD}'_{T,z_{\alpha}}$. By \Cref{eq:small-bias}, for every $T\subseteq [A]\backslash Q$ s.t. $1\le|T|\le t$, 
$$\left(\bigoplus_{\alpha\in T}\bigoplus_{z\in\bits{c}}\ACB(\bX,\Samp(\bX,\alpha,z),(\alpha, z))\right)\approx_{A^{-t}\delta} \bU_1.$$
By \Cref{lemma:AGM} this implies that with probability $1-\delta$ over the fixing of $\bX_2$, 
$$\Reduce(\bX)=\left\{\bigoplus_{z\in\bits{c}}\ACB(\bX,\Samp(\bX,\alpha,z),(\alpha, z))\right\}_{\alpha\in[A]}$$
is $2\delta$-close to a $(2\eps A,t)$-NOBF source. Therefore $\Reduce(\bX)$ is $3\delta$-close to a convex combination of $(2\eps A, t)$-NOBF source.
\end{proof}
To get \Cref{lemma:main}, we need the following affine correlation breaker, which we construct in \Cref{sec:affine-to-std}.  
\begin{theorem}\label{thm:acb}
For every $m,a,t\in\bbN$ and $\eps>0$ there exists an explicit strong $t$-affine correlation breaker $\ACB:\bits{n}\times\bits{d}\times\bits{a}\to\bits{m}$ with error $\eps$ for entropy $k$ such that the seed length is  ${d=O\left(t\log\left(\frac{n}{\eps}\right)\cdot\left(\frac{\log(a)}{\log\log(a)}+\log^3(t)\right)\right)}$ and ${k=O\left(tm+t\log\left(\frac{n}{\eps}\right)\cdot\left(\frac{\log(a)}{\log\log(a)}+t\right)\right)}$.
\end{theorem}
Now we are ready to prove \Cref{lemma:main}.
\begin{proof}[Proof of \Cref{lemma:main}]
Let $\Samp:\bits{n}\times[N]\times[C]\to\bits{d}$ be a $(N^{-\gamma}/2,N^{-\gamma}/3)$-somewhere random sampler from $\Cref{lemma:sr-sampler}$, where $N=n^{O(1)}$. We want to choose proper parameters $d,C$ so that there exists a $(Ct-1)$-strong affine correlation breaker $\ACB:\bits{n}\times\bits{d}\times[NC]\to\bits{}$ with error $N^{-2(t+\gamma)}C^{-1}/6$. Then \Cref{lemma:compose} would imply \Cref{lemma:main}. Observe that we need to guarantee $$d\ge 
K_1\left(Ct^2\log\left(n\right)\cdot\left(\frac{\log\log(n)}{\log\log\log(n)}+\log^3(Ct)\right)\right)$$
and
 $$C\ge K_2 \log^2\left(\frac{d}{\log(n)}\right)$$
for some fixed constants $K_1,K_2$. It suffices to take $C=O(\log\log\log^2(n)+\log^2(t))$ for some large enough constant factor. Then the entropy requirement of $\ACB$ would be 
$$k_1= O\left(Ct^2\log\left(n\right)\cdot\left(\frac{\log\log(n)}{\log\log\log(n)}+Ct\right)\right),$$
and the entropy requirement of $\Samp$ would be $k_2=O(d+\log(N^{\gamma}))=O(d+\log(n))$. To make $\Reduce$ work, the entropy of the given sumset source should be at least 
$$k=\max\{k_1+Ctd,k_2\}=O\left(C^2t^3\log\left(n\right)\cdot\left(\frac{\log\log(n)}{\log\log\log(n)}+\log^3(t)\right)\right).$$
Finally, observe that the running time of $\Reduce$ is $N$ times the running time of $\ACB$ and $\Samp$, which is also $\poly(n)$. 
\end{proof}
\section{Construction of Affine Correlation Breakers}\label{sec:affine-to-std}
In this section we prove \Cref{thm:acb-to-std}, which we restate below. 
\begin{theorem}[\Cref{thm:acb-to-std}, restated]
Let $C$ be a large enough constant. Suppose that there exists an explicit $(d_0,\eps)$-strong correlation breaker $\CB:\bits{d}\times\bits{d_0}\times\bits{a}\to\bits{C\log^2(t+1)\log(n/\eps)}$ for some $n,t\in\bbN$. Then there exists an explicit strong $t$-affine correlation breaker $\ACB:\bits{n}\times\bits{d}\times\bits{a}\to\bits{m}$ with error $O(t\eps)$ for entropy $k=O(td_0+tm+t^2\log(n/\eps))$, where $d=O(td_0+m+t\log^3(t+1)\log(n/\eps))$.
\end{theorem}
We note that it is possible to get different trade-off between the entropy $k$ and the seed length $d$. Here we focus on minimizing $\min(k,td)$, which corresponds to the entropy of our extractors. With \Cref{thm:acb-to-std} we directly get \Cref{thm:acb} by plugging in the following (standard) correlation breaker by Li~\cite{Li19}.
\begin{theorem}[\cite{Li19}]
There exists an explicit (standard) correlation breaker $\bits{n}\times\bits{d}\times\bits{a}\to\bits{m}$ for entropy $d$ with error $\eps$, where $d=O\left(m+\log(n/\eps)\cdot\frac{\log(a)}{\log\log(a)}\right)$.
\end{theorem}

\begin{proof}[Proof of \Cref{thm:acb-to-std}]
Consider any $\bA,\bB\in\bits{n},\bY,\bY^{[t]}\in\bits{d},\bZ\in\bits{*}$ such that 
\begin{itemize}
\item 
$\bA\lra \bZ \lra (\bB,\bY,\bY^{[t]})$ forms a Markov chain
\item
$\avghmin(\bA\mid \bZ)\ge k$
\item
$(\bY,\bZ)=(\bU,\bZ)$,
\end{itemize}
and any $\alpha,\alpha^{[t]}\in\bits{a}$ such that $\alpha\neq \alpha^i$ for every $i\in[t]$. Let $\bX=\bA+\bB$. Our goal is to construct an algorithm $\ACB$ and prove that 
\begin{equation}
(\ACB(\bX,\bY,\alpha)\approx_{O(t\eps)} \bU_m) \mid (\{\ACB(\bX,\bY^i,\alpha^i)\}_{i\in[t]},\bY,\bY^{[t]}).
\end{equation}
For readability, first we explain some conventions in our proof. First we note that whenever we define a new random variable $\bV:=f(\bX,\bA,\bB,\bY)$ using some deterministic function $f$, we also implicitly define $\bV^i:=f(\bX,\bA,\bB,\bY^i)$ for every $i\in[t]$. In each step of the proof, we consider a Markov chain $(\bA,\bR)\lra \bZ' \lra (\bB,\bY,\bY^{[t]},\bS)$ for some random variables $\bR,\bZ',\bS$, where $\bR$ is a deterministic function of $(\bA,\bZ')$, and $\bS$ is a deterministic function of $(\bB,\bY,\bY^{[t]},\bZ')$. Initially $\bZ'=\bZ$. When we say ``$\bR$ is $\eps$-close to uniform" it means $(\bR\approx_\eps \bU)\mid \bZ'$, and similarly ``$\bS$ is $\eps$-close to uniform" means $(\bS\approx_\eps \bU)\mid \bZ'$. When we say $\bR$ is independent of $\bS$ it implicitly means $\bR\lra \bZ'\lra \bS$ is a Markov chain. Then when we say ``fix $f(\bR,\bZ)$" for some deterministic function $f$, we consider the Markov chain $(\bA,\bR)\lra (\bZ',f(\bR,\bZ')) \lra (\bB,\bY,\bY^{[t]},\bS)$ in the next step. Similarly when we say ``fix $g(\bS,\bZ)$" for some deterministic function $g$, we consider the Markov chain $(\bA,\bR)\lra (\bZ',g(\bS,\bZ')) \lra (\bB,\bY,\bY^{[t]},\bS)$ in the next step. To make the notations cleaner, sometimes we only specify a Markov chain $\bR\lra \bZ'\lra \bS$ where $\bR,\bS$ are the random variables used in the current step of argument (e.g. when we apply \Cref{lemma:ind-merging}), but it should always be true that $(\bA,\bR)\lra \bZ' \lra (\bB,\bY,\bY^{[t]},\bS)$ is a Markov chain.

The algorithm $\ACB$ consists of two phases. First, let $r=(t+c_{\ref{lemma:linear-ext}}+10)\cdot c_{\ref{lemma:GUV-ext}}\log(n/\eps)$, and let $\LExt_0:\bits{n}\times\bits{d_0'}\to\bits{d_0}$ and $\LExt_{\sfr}:\bits{n}\times\bits{d_x}\to\bits{r}$ be strong linear seeded extractors in \Cref{lemma:linear-ext} with error $\eps$. It suffices to take $d_0'=O(d_0+\log(n/\eps))$ and $d_x=O(\log^2(t+1)\log(n/\eps))$. Therefore if the constant $C$ in the theorem statement is large enough, we can also take the output length of $\CB$ to be $d_x$. The first phase of $\ACB(\bX,\bY,\alpha)$ consists of the following steps. 
\begin{enumerate}
\item 
Let $\bS_1:=\Prefix(\bY,d_0')$. 
\item
Compute $\bR_1:=\LExt_0(\bX,\bS_1)$.
\item
Compute $\bS_2:=\CB(\bY,\bR_1,\alpha)$.
\item
Output $\bR_2:=\LExt_{\sfr}(\bX,\bS_2)$.
\end{enumerate}
Furthermore, define $\bR_{1,\bA}:=\LExt_1(\bA,\bS_1)$, $\bR_{1,\bB}:=\LExt_1(\bB,\bS_1)$, $\bR_{2,\bA}:=\LExt_2(\bA,\bS_2)$ and $\bR_{2,\bB}=\LExt_2(\bB,\bS_2)$, and let $\bZ_0=(\bZ,\bS_1,\bS_1^{[t]},\bR_{1,\bB},\bR_{1,\bB}^{[t]},\bR_{1},\bR_{1}^{[t]},\bS_2,\bS_2^{[t]})$. First we prove that for every $i\in[t]$,
\begin{equation}\label{eq:first-phase}
(\bR_{2,\bA}\approx_{5\eps} \bU) \mid (\bR_{2,\bA}^{i},\bZ_0,\bR_{2,\bB},\bR_{2,\bB}^{[t]}),
\end{equation}
and 
\begin{equation}\label{eq:first-phase-Markov}
(\bA,\bR_{2,\bA},\bR_{2,\bA}^{[t]})\lra \bZ_0 \lra (\bB,\bR_{2,\bB},\bR_{2,\bB}^{[t]},\bY,\bY^{[t]})\textrm{ forms a Markov chain.}
\end{equation}
Note that this means if we output $\bR_2$ we already get a $1$-affine correlation breaker. To prove (\ref{eq:first-phase}) and (\ref{eq:first-phase-Markov}), first note that by definition of $\LExt_0$, we get $(\bR_{1,\bA}\approx_{\eps} \bU_{d_0})\mid (\bZ,\bS_1,\bS_1^{[t]})$. Fix $(\bS_1,\bS_1^{[t]})$. Since $(\bR_{1,\bA},\bR_{1,\bA}^{[t]})$ are deterministic functions of $(\bA,\bS_1,\bS_1^{[t]})$, and $(\bR_{1,\bB},\bR_{1,\bB}^{[t]})$ are deterministic functions of $(\bB,\bS_1,\bS_1^{[t]})$, $(\bR_{1,\bA},\bR_{1,\bA}^{[t]})$ are independent of $(\bR_{1,\bB},\bR_{1,\bB}^{[t]})$. 
Fix $(\bR_{1,\bB},\bR_{1,\bB}^{[t]})$. Then $\bR_{1,\bA}$ is still close to uniform. Because $\bR_{1}=\bR_{1,\bA}+\bR_{1,\bB}$, this implies $$(\bR_1\approx_{\eps} \bU_{d_0})\mid (\bZ,\bS_1,\bS_1^{[t]},\bR_{1,\bB},\bR_{1,\bB}^{[t]}).$$ 
Moreover, $\avghmin(\bY\mid \bZ,\bS_1,\bS_1^{[t]},\bR_{1,\bB},\bR_{1,\bB}^{[t]})\ge d-O(t(d_0+\log(n/\eps)))\ge d_0+\log(1/\eps)$. 
Because $$(\bR_1,\bR_1^{[t]})\lra (\bZ,\bS_1,\bS_1^{[t]},\bR_{1,\bB},\bR_{1,\bB}^{[t]}) \lra (\bY,\bY^{[t]})$$ is a Markov chain, and because $\CB$ is a strong correlation breaker, for every $i\in[t]$ we have 
$$(\bS_2\approx_{3\eps} \bU_{d_x})\mid (\bS_2^i,\bZ,\bS_1,\bS_1^{[t]},\bR_{1,\bB},\bR_{1,\bB}^{[t]},\bR_{1},\bR_{1}^{i}).$$
Note that after fixing $\bR_1$, $\bS_2$ becomes independent of $\bR_1^{[t]}$. Therefore 
$$(\bS_2\approx_{3\eps} \bU_{d_x})\mid (\bS_2^i,\bZ,\bS_1,\bS_1^{[t]},\bR_{1,\bB},\bR_{1,\bB}^{[t]},\bR_{1},\bR_{1}^{[t]}).$$
Fix $\bR_{1},\bR_{1}^{[t]}$. Because $\bA$ are independent of $\bS_2,\bS_2^{[t]}$, by \Cref{lemma:ind-merging} 
we can conclude that
\begin{equation*}
(\bR_{2,\bA}\approx_{5\eps}\bU_r)\mid (\bR_{2,\bA}^i,\bZ,\bS_1,\bS_1^{[t]},\bR_{1,\bB},\bR_{1,\bB}^{[t]},\bR_{1},\bR_{1}^{[t]},\bS_2,\bS_2^{[t]})
\end{equation*}
which is exactly 
\begin{equation}\label{eq:first-phase-last}
(\bR_{2,\bA}\approx_{5\eps}\bU_r)\mid (\bR_{2,\bA}^i,\bZ_0).
\end{equation}
Finally, fix $\bS_2,\bS_2^{[t]}$. Since $(\bR_{2,\bA},\bR_{2,\bA}^{[t]})$ are independent of $(\bR_{2,\bB},\bR_{2,\bB}^{[t]})$, we get (\ref{eq:first-phase-Markov}). Then because $\bR_2=\bR_{2,\bA}+\bR_{2,\bB}$, by (\ref{eq:first-phase-Markov}) and (\ref{eq:first-phase-last}) we get (\ref{eq:first-phase}).

Next we move to the second phase. Let $d_y=c_{\ref{lemma:GUV-ext}}\log(n/\eps)$. Moreover, let $\Ext:\bits{d}\times\bits{d_y}\to\bits{d_x}$ be a strong seeded extractor from \Cref{lemma:GUV-ext}, and $\LExt_{\sfm}:\bits{r}\times\bits{d_x}\to\bits{d_y}$ be a linear strong seeded extractor from \Cref{lemma:linear-ext}.
Define $\bW_{0,\bA}:=\bR_{2,\bA},\bW_{0,\bB}:=\bR_{2,\bB},\bW_0:=\bR_2$ and $h=\ceil{\log t}$. Then repeat the following steps for $i$ from $1$ to $h$:
\begin{enumerate}
\item
Let $\bW_{\sfp,i-1}:=\Prefix(\bW_{i-1},d_y)$.
\item
Compute $\bQ_{\sfm,i-1}:=\Ext(\bY,\bW_{\sfp,i-1})$.
\item
Compute $\bV_i:=\LExt_{\sfm}(\bW_{i-1},\bQ_{\sfm,i-1})$.
\item
Compute $\bQ_{\sfr,i}:=\Ext(\bY,\bV_{i})$.
\item
Compute $\bW_{i}:=\LExt_{\sfr}(\bX,\bQ_{\mathsf{r},i})$.
\end{enumerate}
Note that Step $1- 3$ are the ``independence merging" steps, which computes $\bV_i$ that is independent of every $2^{i}$ tampered versions. Since the length of $\bV_i$ is shorter than $\bW_i$, we use Step $4- 5$ to recover the length and get $\bW_i$ s.t. $|\bW_i|=r$. We claim that each of $\bW_i,\bQ_{\sfm,i},\bV_i,\bQ_{\sfr,i}$ is independent of every $\min(2^i,t)$ tampered versions, and in particular $(\bW_h,\bW_h^{[t]})\approx (\bU_r,\bW_h^{[t]})$. 

Formally, for every $i$ from $1$ to $h$, let $\bW_{\sfp,i-1,\bA}:=\Prefix(\bW_{i-1,\bA},d_y)$, $\bW_{\sfp,i-1,\bB}:=\Prefix(\bW_{i-1,\bB},d_y)$, $\bV_{i,\bA}:=\LExt_{\sfm}(\bW_{i-1,\bA},\bQ_{\sfm,i-1})$, $\bV_{i,\bB}:=\LExt_{\sfm}(\bW_{i-1,\bB},\bQ_{\sfm,i-1})$, $\bW_{i,\bA}:=\LExt_{\sfr}(\bA,\bQ_{\mathsf{r},i})$ and $\bW_{i,\bB}:=\LExt_{\sfr}(\bB,\bQ_{\mathsf{r},i})$. Moreover, for every $i\in[h]$, let 
$$\bZ_i:=\left(\bZ_{i-1},\bW_{\sfp,i-1,\bB},\bW_{\sfp,i-1,\bB}^{[t]},\bW_{\sfp,i-1},\bW_{\sfp,i-1}^{[t]},\bQ_{\sfm,i-1},\bQ_{\sfm,i-1}^{[t]},\bV_{i,\bB},\bV_{i,\bB}^{[t]},\bV_{i},\bV_{i}^{[t]},\bQ_{\sfr,i},\bQ_{\sfr,i}^{[t]}\right).$$
We want to prove the following claims for every $i \in [h]$ by induction:
\begin{itemize}
\item 
For every $T\subseteq [t]$ s.t. $|T|=2^{i}$, 
\begin{equation}\label{eq:second-phase}
(\bW_{i,\bA}\approx_{(13\cdot 2^{i}-8)\eps} \bU_r) \mid (\bW_{i,\bA}^T,\bZ_i).
\end{equation}
\item
The following is a Markov chain:
\begin{equation}\label{eq:second-phase-Markov}
(\bA,\bW_{i,\bA},\bW_{i,\bA}^{[t]})\lra \bZ_i \lra (\bB,\bW_{i,\bB},\bW_{i,\bB}^{[t]},\bY,\bY^{[t]}).
\end{equation}
\end{itemize}
Note that by (\ref{eq:first-phase}) and (\ref{eq:first-phase-Markov}), the conditions above hold for $i=0$. Now assume by induction that (\ref{eq:second-phase}) and (\ref{eq:second-phase-Markov}) hold for $i-1$, and we want to prove (\ref{eq:second-phase}) and (\ref{eq:second-phase-Markov}) for $i$. First, observe that because $\bW_{\sfp,i-1}=\bW_{\sfp,i-1,\bA}+\bW_{\sfp,i-1,\bB}$, by (\ref{eq:second-phase}) and (\ref{eq:second-phase-Markov}) for every $T_1\subseteq [t]$ of size $2^{i-1}$, 
$$(\bW_{\sfp,i-1}\approx_{(13\cdot 2^{i-1}-8)\eps} \bU_r) \mid (\bW_{\sfp,i-1}^{T_1},\bZ_{i-1},\bW_{\sfp,i-1,\bB},\bW_{\sfp,i-1,\bB}^{[t]}).$$ Fix $(\bW_{\sfp,i-1,\bB},\bW_{\sfp,i-1,\bB}^{[t]})$. Note that 
$$(\bW_{\sfp,i-1},\bW_{\sfp,i-1}^{[t]})\lra (\bZ_{i-1},\bW_{\sfp,i-1,\bB},\bW_{\sfp,i-1,\bB}^{[t]}) \lra (\bY,\bY^{[t]})$$
is a Markov chain.
By \Cref{lemma:ind-merging} (similarly we omit the entropy requirement for $\bY$ for now and will verify it in the end), for every $T_1\subseteq [t]$ of size $2^{i-1}$, 
$$(\bQ_{\sfm,i-1}\approx_{(13\cdot 2^{i-1}-6)\eps} \bU_{d_x}) \mid (\bQ_{\sfm,i-1}^{T_1},\bZ_{i-1},\bW_{\sfp,i-1,\bB},\bW_{\sfp,i-1,\bB}^{[t]},\bW_{\sfp,i-1},\bW_{\sfp,i-1}^{[t]}).$$
Next, fix $(\bW_{\sfp,i-1},\bW_{\sfp,i-1}^{[t]})$. Now consider any $T\subseteq [t]$ s.t. $|T|=\min(2^i,t)$, and any $T_1,T_2$ s.t. $|T_1|=|T_2|=2^{i-1}$ and $T_1\cup T_2=T$. By (\ref{eq:second-phase}) there exists $\bW_{i-1,\bA}'=\bU_r$ s.t. $$(\bW_{i-1,\bA}\approx_{(13\cdot 2^{i-1}-8)\eps} \bW_{i-1,\bA}')\mid (\bW_{i-1,\bA}^{T_2},\bZ_{i-1},\bW_{\sfp,i-1,\bB},\bW_{\sfp,i-1,\bB}^{[t]},\bW_{\sfp,i-1},\bW_{\sfp,i-1}^{[t]})$$
and 
$$\avghmin\left( \bW_{i-1,\bA}'\mid \bW_{i-1,\bA}^{T_2},\bZ_{i-1},\bW_{\sfp,i-1,\bB},\bW_{\sfp,i-1,\bB}^{[t]},\bW_{\sfp,i-1},\bW_{\sfp,i-1}^{[t]}\right)\ge r-(t+1)d_y.$$
Let $\bZ_{i-1}':=\left(\bZ_{i-1},\bW_{\sfp,i-1,\bB},\bW_{\sfp,i-1,\bB}^{[t]},\bW_{\sfp,i-1},\bW_{\sfp,i-1}^{[t]},\bQ_{\sfm,i-1},\bQ_{\sfm,i-1}^{[t]}\right)$. By \Cref{lemma:ind-merging},
$$(\bV_{i,\bA}\approx_{(13\cdot 2^{i}-14)\eps} \bU_{d_y}) \mid \left(\bV_{i,\bA}^T, \bZ_{i-1}' \right).$$
Fix $(\bQ_{\sfm,i-1},\bQ_{\sfm,i-1}^{[t]})$. Note that $\bZ_{i-1}'$ consists of exactly the random variables we have fixed so far. Because $\bV_{i}=\bV_{i,\bA}+\bV_{i,\bB}$ and $(\bV_{i,\bA},\bV_{i,\bA}^{[t]})\lra \bZ_{i-1}' \lra (\bV_{i,\bB},\bV_{i,\bB}^{[t]})$ forms a Markov chain,
$$(\bV_{i}\approx_{(13\cdot 2^{i}-12)\eps} \bU_{d_y}) \mid \left(\bV_{i}^T, \bZ_{i-1}', \bV_{i,\bB},\bV_{i,\bB}^{[t]}\right).$$
Next we fix $(\bV_{i,\bB},\bV_{i,\bB}^{[t]})$. Since $(\bV_{i},\bV_{i}^{[t]})\lra (\bZ_{i-1}', \bV_{i,\bB},\bV_{i,\bB}^{[t]}) \lra (\bY,\bY^{[t]})$, again by \Cref{lemma:ind-merging},
$$(\bQ_{\sfr,i}\approx_{(13\cdot 2^{i}-10)\eps} \bU_{d_x}) \mid \left(\bQ_{\sfr,i}^T, \bZ_{i-1}', \bV_{i,\bB},\bV_{i,\bB}^{[t]},\bV_{i},\bV_{i}^{[t]}\right).$$
Next, fix $(\bV_{i},\bV_{i}^{[t]})$. Since $\bA \lra (\bZ_{i-1}', \bV_{i,\bB},\bV_{i,\bB}^{[t]},\bV_{i},\bV_{i}^{[t]})\lra (\bQ_{\sfr,i},\bQ_{\sfr,i}^{[t]})$, by \Cref{lemma:ind-merging}
$$(\bW_{i,\bA}\approx_{(13\cdot 2^{i}-8)\eps} \bU_r)\mid \left(\bZ_{i-1}', \bV_{i,\bB},\bV_{i,\bB}^{[t]},\bV_{i},\bV_{i}^{[t]},\bQ_{\sfr,i},\bQ_{\sfr,i}^{[t]}\right),$$
which is exactly (\ref{eq:second-phase}). Fix $(\bQ_{\sfr,i},\bQ_{\sfr,i}^{[t]})$. Because $(\bW_{i,\bA},\bW_{i,\bA}^{[t]})$ are deterministic functions of $(\bA,\bQ_{\sfr,i},\bQ_{\sfr,i}^{[t]})$ and $(\bW_{i,\bB},\bW_{i,\bB}^{[t]})$ are deterministic functions of $(\bB,\bQ_{\sfr,i},\bQ_{\sfr,i}^{[t]})$, we get (\ref{eq:second-phase-Markov}). Finally we need to verify that whenever we apply \Cref{lemma:ind-merging}, $\bX$ and $\bY$ have enough conditional entropy. Observe that every time we apply \Cref{lemma:ind-merging} on $\bA$, we condition on some random variables in $\bZ_h$, take an extractor from \Cref{lemma:linear-ext} with error $\eps$ and output at most $r$ bits. The conditional entropy of $\bA$ is at least 
$$\avghmin(\bA\mid \bZ_h)\ge \avghmin(\bA\mid \bZ) - (t+1)\cdot O(d_0+\log(n/\eps)+h(d_x+d_y)) \ge (t+c_{\ref{lemma:linear-ext}})r + \log(1/\eps),$$
which satisfies the requirement in \Cref{lemma:ind-merging}.
Every time we apply \Cref{lemma:ind-merging} on $\bY$, we condition on some random variables in $\bZ_h$, take an extractor from \Cref{lemma:GUV-ext} with error $\eps$ and output at most $d_x$ bits. The conditional entropy of $\bY$ is at least 
$$\avghmin(\bY\mid \bZ_h)\ge d - (t+1)\cdot O(d_0+\log(n/\eps)+h(d_x+d_y)) \ge (t+2)d_x + \log(1/\eps),$$
which satisfies the requirement in \Cref{lemma:ind-merging}.

Since $\bW_h=\bW_{h,\bA}+\bW_{h,\bB}$, (\ref{eq:second-phase}) and (\ref{eq:second-phase-Markov}) together imply 
$$(\bW_{h}\approx_{(13t-8)\eps} \bU_r)\mid (\bW_{h}^{[t]},\bY,\bY^{[t]}).$$ Therefore if $m\le r$, it suffices to output $\ACB(\bX,\bY,\alpha)=\Prefix(\bW_{h},m)$. If $m>r$, we can do one more round of alternating extraction to increase the output length. Let $\LExt_{\mathsf{out}}:\bits{n}\times\bits{d_{\mathsf{out}}}\to\bits{m}$ be a linear strong seeded extractor with error $\eps$ from \Cref{lemma:linear-ext} and $\Ext_{\mathsf{out}}:\bits{d}\times\bits{r}\to\bits{d_{\mathsf{out}}}$ be a seeded extractor from \Cref{lemma:GUV-ext}. It suffices to take $d_{\mathsf{out}}=O\left(\frac{m}{t}+\log^2(t+1)\log(\frac{n}{\eps})\right)$. Then 
\begin{enumerate}
\item 
Compute $\bQ_{\mathsf{out}}:=\Ext_{\mathsf{out}}(\bY,\bW_h)$.
\item
Output $\bW_{\mathsf{out}}:=\LExt_{\mathsf{out}}(\bX,\bQ_{\mathsf{out}})$.
\end{enumerate}
Since $(\bW_h\approx\bU)\mid (\bZ_i,\bW_{h,\bB},\bW_{h,\bB}^{[t]})$, $(\bW_h,\bW_h^{[t]})\lra (\bZ_i,\bW_{h,\bB},\bW_{h,\bB}^{[t]})\lra (\bY,\bY^{[t]})$ forms a Markov chain and 
$$\avghmin(\bY\mid \bZ_i,\bW_{h,\bB},\bW_{h,\bB}^{[t]})\ge d-(t+1)\cdot O(d_0+\log(n/\eps)+h(d_x+d_y))\ge (t+2)d_{\mathsf{out}}+\log(1/\eps),$$
by \Cref{lemma:ind-merging} 
$$(\bQ_{\mathsf{out}}\approx_{(13t-6)\eps}  \bU_{d_{\mathsf{out}}}) \mid (\bQ_{\mathsf{out}}^{[t]},\bZ_i,\bW_{h,\bB}, \bW_{h,\bB}^{[t]}, \bW_{h},\bW_{h}^{[t]}).$$
And because $\bA$ is independent of $(\bQ_{\mathsf{out}},\bQ_{\mathsf{out}}^{[t]})$ conditioned on $(\bZ_i,\bW_{h,\bB}, \bW_{h,\bB}^{[t]}, \bW_{h},\bW_{h}^{[t]})$, and 
$$\avghmin(\bA\mid \bZ_i,\bW_{h,\bB},\bW_{h,\bB}^{[t]},\bW_{h},\bW_{h}^{[t]})\ge k-(t+1)\cdot O(d_0+\log(n/\eps)+h(d_x+d_y))\ge (t+2)d_{\mathsf{out}}+\log(1/\eps),$$
again by \Cref{lemma:ind-merging} we can conclude that 
$$(\bW_{\mathsf{out},\bA}\approx_{(13t-4)\eps} \bU_m)\mid (\bW_{\mathsf{out},\bA}^{[t]},\bZ_i,\bW_{h,\bB},\bW_{h,\bB}^{[t]},\bW_{h},\bW_{h}^{[t]},\bQ_{\mathsf{out}},\bQ_{\mathsf{out}}^{[t]}).$$
Since $\bW_{\mathsf{out}}=\bW_{\mathsf{out},\bA}+\bW_{\mathsf{out},\bB}$ and $(\bW_{\mathsf{out},\bA},\bW_{\mathsf{out},\bA}^{[t]})$ are independent of $(\bY,\bY^{[t]},\bW_{\mathsf{out},\bA},\bW_{\mathsf{out},\bA}^{[t]})$ conditioned on $(\bZ_i,\bW_{h,\bB},\bW_{h,\bB}^{[t]},\bW_{h},\bW_{h}^{[t]},\bQ_{\mathsf{out}},\bQ_{\mathsf{out}}^{[t]})$, we can conclude that 
$$(\bW_{\mathsf{out}}\approx_{(13t-4)\eps} \bU_m)\mid (\bW_{\mathsf{out}}^{[t]},\bZ,\bY,\bY^{[t]}),$$
which means $\ACB(\bX,\bY,\alpha)=\bW_{\mathsf{out}}$ is a strong $t$-affine correlation breaker with error $O(t\eps)$.

\end{proof}
\section{Sumset Sources with Small Doubling}\label{sec:small-doubling}
In this section we show that a sumset source with small doubling constant is close to a convex combination of affine sources, as stated in \Cref{thm:affine-sumset}. To prove this result, first we need \Cref{lemma:Croot-Sisask}, which is a variant of the Croot-Sisask lemma~\cite{CS10}. For the proof of \Cref{lemma:Croot-Sisask} we follow the exposition by Ben-Sasson, Ron-Zewi, Tulsiani and Wolf \cite{BRTW14} which is more convenient for our setting.
\begin{lemma}\label{lemma:Croot-Sisask}
Let $A\subseteq \bbF_2^n$ be a set which satisfies $\abs{A}\ge \abs{\bbF_2^n}/r$. Then for every $\eps>0$ and every pair of functions $f,g:\bbF_2^n\to{[0,1]}$ there exists $t=O(\log(r/\eps)/\eps^2)$ and a set $X$ of size at least $\abs{\bbF_2^n}/2r^t$ such that for every set $B$ s.t. $\abs{B}\ge \abs{\bbF_2^n}/r$ and every $x\in X$, 
$$\ex[a\sim A, b\sim B]{f(a+b)}\approx_{\eps} \ex[a\sim A, b\sim B]{f(a+b+x)}$$
and 
$$\ex[a\sim A, b\sim B]{g(a+b)}\approx_{\eps} \ex[a\sim A, b\sim B]{g(a+b+x)}.$$
\end{lemma}
\begin{proof}
Let $t=8\ln(128r/\eps)/\eps^2$. By Chernoff-Hoeffding bound, for every $b\in \bbF_2^n$,
$$\pr[(a_1,\dots,a_t)\sim A^t]{\frac{1}{t}\sum_{i=1}^t f(a_i+b) \approx_{\frac{\eps}{4}} \ex[a\sim A]{f(a+b)} }\ge 1-\frac{\eps}{16r}$$
and 
$$\pr[(a_1,\dots,a_t)\sim A^t]{\frac{1}{t}\sum_{i=1}^t g(a_i+b) \approx_{\frac{\eps}{4}} \ex[a\sim A]{g(a+b)} }\ge 1-\frac{\eps}{16r}.$$
Then by union bound and by averaging over $b\sim\bbF_2^n$,
$$\pr[\substack{(a_1,\dots,a_t)\sim A^t \\ b\sim \bbF_2^n}]{\frac{1}{t}\sum_{i=1}^t f(a_i+b) \approx_{\frac{\eps}{4}} \ex[a\sim A]{f(a+b)} \textrm{ and }\frac{1}{t}\sum_{i=1}^t g(a_i+b) \approx_{\frac{\eps}{4}} \ex[a\sim A]{g(a+b)}}\ge 1-\frac{\eps}{8r}.$$
Define $$\mathsf{BAD}_{(a_1,\dots,a_t)}:=\left\{b: \frac{1}{t}\sum_{i=1}^t f(a_i+b) \not\approx_{\frac{\eps}{4}} \ex[a\sim A]{f(a+b)} \textrm{ or }\frac{1}{t}\sum_{i=1}^t g(a_i+b) \not\approx_{\frac{\eps}{4}} \ex[a\sim A]{g(a+b)}\right\}.$$
By Markov inequality, there exists $S\subseteq A^t$ such that $\abs{S}\ge \abs{A}^t/2$ and for every $(a_1,\dots,a_t)\in S$,
$$\abs{\mathsf{BAD}_{(a_1,\dots,a_t)}}\le \frac{\eps}{4r}\abs{\bbF_2^n}.$$

Now classify the elements in $S$ by $(a_2-a_1,a_3-a_1,\dots,a_t-a_1)$. By averaging there exists a subset $X'\subseteq S$ and a $(t-1)$-tuple $(y_2,\dots,y_{t})$ such that $\abs{X'}\ge \abs{S}/\abs{\bbF_2^n}^{t-1}\ge \abs{\bbF_2^n}/2r^t$, and for every $(a_1,\dots,a_t)\in X'$ we have $a_{i}-a_1=y_i$ for every $2\le i\le t$. Let $(a_1^*,\dots,a_t^*)$ be an element in $X'$. Observe that for every $(a_1,\dots,a_t)\in X'$, $a_1-a_1^*=\dots=a_t-a_t^*$. Define 
$$X=\{x=a_1-a_1^*: (a_1,\dots,a_t)\in X'\}.$$
Note that $\abs{X}=\abs{X'}\ge \abs{\bbF_2^n}/2r^t$. It remains to prove that for every $x\in X$, 
$$\ex[a\sim A, b\sim B]{f(a+b)}\approx_{\eps} \ex[a\sim A, b\sim B]{f(a+b+x)}$$
and 
$$\ex[a\sim A, b\sim B]{g(a+b)}\approx_{\eps} \ex[a\sim A, b\sim B]{g(a+b+x)}.$$
Let $(a_1,\dots,a_t)=(a_1^*+x,\dots,a_t^*+x)$. Since $(a_1,\dots,a_t)$ is an element in $S$, 
$$\abs{\ex[a\sim A, b\sim B]{f(a+b)}- \ex[b\sim B]{\frac{1}{t}\sum_{i=1}^t f(a_i+b)}}\le \frac{\eps}{4}+ \pr[b\sim B]{b\in \mathsf{BAD}_{(a_1,\dots,a_t)}}\le \frac{\eps}{2}.$$
Similarly, since $(a_1^*,\dots,a_t^*)$ is an element in $S$,
$$\abs{\ex[a\sim A, b\sim B]{f(a+b+x)}- \ex[b\sim B]{\frac{1}{t}\sum_{i=1}^t f(a_i^*+b+x)}}\le \frac{\eps}{4}+ \pr[b\sim B]{(b+x)\in \mathsf{BAD}_{(a_1^*,\dots,a_t^*)}}\le \frac{\eps}{2}.$$
Finally, observe that 
$$ \ex[b\sim B]{\frac{1}{t}\sum_{i=1}^t f(a_i+b)}= \ex[b\sim B]{\frac{1}{t}\sum_{i=1}^t f(a_i^*+x+b)}.$$
By triangle inequality we can conclude that 
$$\ex[a\sim A, b\sim B]{f(a+b)}\approx_{\eps} \ex[a\sim A, b\sim B]{f(a+b+x)}.$$ 
Similarly we can prove that 
$$\ex[a\sim A, b\sim B]{g(a+b)}\approx_{\eps} \ex[a\sim A, b\sim B]{g(a+b+x)}.$$
\end{proof}
Next we prove the following lemma. The proof is along the lines of \cite[Theorem A.1]{Sanders12}. (See also the survey by Lovett~\cite{Lovett15}.)
\begin{lemma}\label{lemma:Sanders}
Let $A,B\subseteq \bbF_2^n$ be sets which satisfy $\abs{A},\abs{B}\ge \abs{\bbF_2^n}/r$. Let $\bA,\bB$ be the uniform distributions over $A,B$ respectively. Then for every $\eps>0$ and every pair of functions $f,g:\bbF_2^n\to{[0,1]}$ there exists a linear subspace $V$ of co-dimension $O(\log^3(r/\eps)\log(r)/\eps^2)$ and a distribution $\bT\in\bbF_2^n$ such that 
$$\ex{f(\bA+\bB)}\approx_{\eps} \ex{f(\bT+\bV)}$$
and 
$$\ex{g(\bA+\bB)}\approx_{\eps} \ex{g(\bT+\bV)},$$
where $\bV$ is the uniform distribution over $V$. 
\end{lemma}
To prove \Cref{lemma:Sanders}, first we need the following corollary of \Cref{lemma:Croot-Sisask}.
\begin{corollary}\label{coro:Croot-Sisask}
Let $A\subseteq \bbF_2^n$ be a set which satisfies $\abs{A}\ge \abs{\bbF_2^n}/r$. Then for every $\eps>0$ and every pair of functions $f,g:\bbF_2^n\to{[0,1]}$ there exists $t=O(\log(r/\eps)/\eps^2)$ and a a set $X$ of size at least $\abs{\bbF_2^n}/2r^t$ such that for every set $B$ s.t. $\abs{B}\ge \abs{\bbF_2^n}/r$ and every $(x_1,\dots,x_\ell )\in X^{\ell}$, 
$$\ex[a\sim A, b\sim B]{f(a+b)}\approx_{\ell\eps} \ex[a\sim A, b\sim B]{f(a+b+x_1+\dots+x_\ell)}$$
and 
$$\ex[a\sim A, b\sim B]{g(a+b)}\approx_{\ell\eps} \ex[a\sim A, b\sim B]{g(a+b+x_1+\dots+x_\ell)}.$$
\end{corollary}
\begin{proof}
Assume by induction that $$\ex[a\sim A, b\sim B]{f(a+b)}\approx_{(\ell-1)\eps} \ex[a\sim A, b\sim B]{f(a+b+x_1+\dots+x_{\ell-1})}.$$
Since $\abs{B+x_1+\dots+x_{\ell-1}}=\abs{B}\ge \abs{\bbF_2^n}/r$, by \Cref{lemma:Croot-Sisask} we get $$\ex[a\sim A, b\sim B]{f(a+b+x_1+\dots+x_{\ell-1})}\approx_{\eps} \ex[a\sim A, b\sim B]{f(a+b+x_1+\dots+x_{\ell})}.$$
Then the claim follows by triangle inequality. The proof for the case of $g$ is exactly the same.
\end{proof}
\begin{proof}[Proof of \Cref{lemma:Sanders}]
Define $\ell=\log(2r/\eps)$. By \Cref{coro:Croot-Sisask} there exists $t=O(\ell^3/\eps^2)$ and a set $X$ of size $\abs{\bbF_2^n}/2r^{t}$ s.t. for every $(x_1,x_2,\dots,x_\ell)\in X^\ell$, 
\begin{equation}\label{eq:period}
\ex{f(\bA+\bB)}\approx_{\eps/2}\ex{f(\bA+\bB+x_1+\dots+x_\ell)}
\end{equation}
and 
$$\ex{g(\bA+\bB)}\approx_{\eps/2}\ex{g(\bA+\bB+x_1+\dots+x_\ell)}.$$
Let $\bX_1,\dots,\bX_\ell$ be independent uniform distributions over $X$. Let $V=\Spec_{1/2}(X)^{\bot}$ and $\bV$ be uniform distribution over $V$. Note that by Chang's lemma (\Cref{lemma:Chang}), $V$ has dimension at least $k'=m-O(\log(r)\log^3(r/\eps)/\eps^2)\ge k-O(\log(r)\log^3(r/\eps)/\eps^2)$. By \Cref{lemma:convolution}, \ref{lemma:expectation} and \ref{lemma:Parseval},
$$\ex{f(\bA+\bB+\bX_1+\dots+\bX_\ell)}=\sum_{\alpha\in\bbF_2^n}\widehat{\mu_A}(\alpha)\widehat{\mu_B}(\alpha)(\widehat{\mu_X}(\alpha))^{\ell}\widehat{f}(\alpha)$$
and 
$$\ex{f(\bA+\bB+\bX_1+\dots+\bX_\ell+\bV)}=\sum_{\alpha\in\bbF_2^n}\widehat{\mu_A}(\alpha)\widehat{\mu_B}(\alpha)(\widehat{\mu_X}(\alpha))^{\ell}\widehat{\mu_V}(\alpha)\widehat{f}(\alpha).$$
Define $\bT=\bA+\bB+\bX_1+\dots+\bX_\ell$. Then
\begin{align*}
\abs{\ex{f(\bT)}-\ex{f(\bT+\bV)}} 
&=\abs{\sum_{\alpha\not\in V^{\bot}}\widehat{\mu_A}(\alpha)\widehat{\mu_B}(\alpha)(\widehat{\mu_X}(\alpha))^{\ell}\widehat{f}(\alpha)} \textrm{ (by \Cref{lemma:density-subspace})}\\
&\le 2^{-\ell}\sum_{\alpha\not\in V^{\bot}}\abs{\widehat{\mu_A}(\alpha)\widehat{\mu_B}(\alpha)\widehat{f}(\alpha)}\textrm{ (by definition of $\Spec_{1/2}(X)$)}\\
&\le 2^{-\ell}\sum_{\alpha\not\in V^{\bot}}\abs{\widehat{\mu_A}(\alpha)\widehat{\mu_B}(\alpha)}\textrm{ (since $\abs{\widehat{f}(\alpha)}\le 1$)}\\
&\le 2^{-\ell} \cdot \sqrt{\left(\sum_{\alpha\in\bbF_2^n}\widehat{\mu_A}(\alpha)^2\right) \left(\sum_{\alpha\in\bbF_2^n}\widehat{\mu_B}(\alpha)^2\right)}\textrm{ (by Cauchy-Schwarz)}\\
&\le 2^{-\ell}\cdot r = \eps/2. \textrm{ (by Parseval's identity (\Cref{lemma:Parseval}))}
\end{align*}
By triangle inequality and (\ref{eq:period}) we get $\ex{f(\bA+\bB)}\approx_{\eps}\ex{f(\bT+\bV)}$. The exact same proof can also show that  $\ex{g(\bA+\bB)}\approx_{\eps}\ex{g(\bT+\bV)}$. 
\end{proof}
Finally, to prove \Cref{thm:affine-sumset}, we need the following lemma.
\begin{lemma}\label{lemma:Freiman-inverse}
Let $X\subseteq \bbF_2^n$ be a set, $\phi:\bbF_2^n\to\bbF_2^m$ be a linear Freiman 3-homorphism of $X$, and $\phi^{-1}:\bbF_2^m\to\bbF_2^n$ be a inverse of $\phi$ such that $\phi^{-1}(\phi(x))=x$ for every $x\in 3X$. (Such a $\phi^{-1}$ exists because $\phi$ is injective on $3X$.) Then for every affine subspace $V\subseteq \bbF_2^m$ such that $\abs{V\cap \phi(X)}> \abs{V}/2$, $\phi^{-1}$ is injective on $V$ and $\phi^{-1}(V)\subseteq \bbF_2^n$ is also an affine subspace.
\end{lemma}
\begin{proof}
Let $t$ be an element in $X$ such that $\phi(t)\in V$. Note that $t$ must exist because $V\cap \phi(X)$ is non-empty. Since $V$ is an affine subspace, for every $v\in V$ and $v_1\in V\cap \phi(X)$, $v+\phi(t)-v_1 \in V$. Because $\abs{V\cap \phi(X)}>\abs{V\backslash \phi(X)}$, for every $v\in V$ there must exist $v_1 \in V\cap \phi(X)$ s.t. $v+\phi(t)-v_1\in V\cap \phi(X)$. In other words, for every $v\in V$ there exist $v_1,v_2\in V\cap \phi(X)$ such that $v=v_1+v_2-\phi(t)$. This means $V\subseteq \phi(2X-t)\subseteq \phi(3X)$.  Because $\phi^{-1}$ is injective on $\phi(3X)$, this implies that $\phi^{-1}$ is injective on $V$. Next we prove that $\phi^{-1}(V)$ is also an affine subspace. 
It suffices to prove that for every $u,v\in V$, $$\phi^{-1}(u)+\phi^{-1}(v)-t=\phi^{-1}(u+v-\phi(t)),$$
because $\phi^{-1}(u+v-\phi(t))\in\phi^{-1}(V)$.
Observe that $$\phi(\phi^{-1}(u)+\phi^{-1}(v)-t-\phi^{-1}(u+v-\phi(t)))=u+v-\phi(t)-(u+v-\phi(t))=0,$$
because $\phi$ is linear, and for every $y\in\{u,v,u+v-\phi(t)\}$ we have $y\in V\subseteq \phi(3X)$, which means $\phi(\phi^{-1}(y))=y$. Moreover, because $\phi^{-1}(u),\phi^{-1}(v),\phi^{-1}(u+v-\phi(t))\in \phi^{-1}(V)\subseteq 2X-t$, 
$$\phi^{-1}(u)+\phi^{-1}(v)-t-\phi^{-1}(u+v-\phi(t))\in 6X.$$
By \Cref{lemma:linear-Freiman}, $\phi^{-1}(u)+\phi^{-1}(v)-t-\phi^{-1}(u+v-\phi(t))=0$.
\end{proof}
Now we are ready to prove \Cref{thm:affine-sumset}.
\begin{proof}[Proof of \Cref{thm:affine-sumset}]
Consider any function $f:\bbF_2^n\to {[0,1]}$. Let $\phi:\bbF_2^n\to\bbF_2^m$ be the $3$-Freiman homomorphism of $A+B$ guaranteed in \Cref{lemma:Green-Rusza}, and let $\phi^{-1}:\bbF_2^n\to\bbF_2^m$ be a inverse of $\phi$ such that $\phi^{-1}(\phi(x))=x$ for every $x\in 3A+3B$. By $\Cref{lemma:linear-Freiman}$, $\phi$ is injective on $A$ and $B$ since $A\subseteq 3(A+B)+b$ for any $b\in B$ and $B\subseteq 3(A+B)+a$ for any $a\in A$. Let $A'=\phi(A),B'=\phi(B),\bA'=\phi(\bA),\bB'=\phi(\bB)$. Observe that $\bA',\bB'$ are exactly the uniform distributions over $A',B'$ respectively. By \Cref{lemma:Green-Rusza} and \Cref{lemma:Plunnecke}, we get $\abs{\bbF_2^m}=\abs{\phi(6A+6B)}\le \abs{6A+6B}\le r^{13}\abs{A}$, which implies $\abs{A'}=\abs{B'}=\abs{A}\ge \abs{\bbF_2^m}/r^{13}$. By \Cref{lemma:Sanders}, there exists a distribution $\bT\in\bbF_2^m$ and a linear subspace $V$ of entropy $k'=m-O(\log(r)\log(r/\eps)^3/\eps^2)$ such that
$$\ex{\mathds{1}_{A'+B'}(\bA'+\bB')}\approx_{\eps/3} \ex{\mathds{1}_{A'+B'}(\bT+\bV)}$$
and 
\begin{equation}\label{eq:f-indistinguishable}
\ex{f(\phi^{-1}(\bA'+\bB'))}\approx_{\eps/3}\ex{f(\phi^{-1}(\bT+\bV))},
\end{equation}
where $\bV$ is the uniform distribution over $V$. Now observe that since $\ex{\mathds{1}_{A'+B'}(\bA'+\bB')}=1$, 
$$\ex{\mathds{1}_{A'+B'}(\bT+\bV)}\ge 1-\eps/3.$$
By Markov's inequality,
$$\pr[t\sim\bT]{\ex{\mathds{1}_{A'+B'}(t+\bV)}>1/2}\ge 1-2\eps/3.$$
In other words, 
$$\pr[t\sim\bT]{\abs{\phi(A+B)\cap(t+V)}>\frac{1}{2}\abs{t+V}}\ge 1-2\eps/3.$$
By \Cref{lemma:Freiman-inverse},
$$\pr[t\sim\bT]{\phi^{-1}(t+\bV)\textrm{ is an affine source of entropy } k'}\ge 1-2\eps/3.$$
Therefore $\phi^{-1}(\bT+\bV)$ is $(2\eps/3)$-close to a convex combination of affine sources (denoted by $\bW$) of entropy $k'$. Since $A+B\subseteq 3A+3B$, $\phi^{-1}(\bA'+\bB')=\phi^{-1}(\phi(\bA+\bB))$ is exactly $\bA+\bB$. Therefore by (\ref{eq:f-indistinguishable}) and triangle inequality, 
$$\ex{f(\bA+\bB)}\approx_\eps \ex{f(\bW)}.$$
Since the proof above works for every function $f:\bbF_2^n\to {[0,1]}$, by \Cref{coro:distribution-minimax}, $\bA+\bB$ is $\eps$-close to a convex combination of affine sources.
\end{proof}

\bibliographystyle{alpha}
\bibliography{extractor}

\appendix 
\section{Proof of \Cref{lemma:BDT}}\label{appendix:BDT-proof}
To prove \Cref{lemma:BDT}, we need the following disperser by Zuckerman~\cite{Zuc07}.
\begin{definition}
We say a function $\Gamma:[N]\times [D]\to {[M]}$ is a \emph{$(K,\eps)$-disperser} if for every set $X\subseteq [N]$ with $|X|\ge K$, the set $\Gamma(X):=\{\Gamma(x,y)\mid x\in X, y\in [D]\}$ satisfies 
$$\abs{\Gamma(X)}\ge \eps M.$$
\end{definition}
\begin{lemma}[\cite{Zuc07}]\label{lemma:Zuc-disperser}
For every constant $\gamma>0$ and $\eps=\eps(n)>0$, there exists an efficient family of $(K=N^{\gamma},\eps)$-disperser $\Gamma:[N=2^n]\times[D]\to{[M]}$ such that $D=O(\frac{n}{\log(1/\eps)})$ and $M=\sqrt{K}$.
\end{lemma}
\begin{proof}[Proof of \Cref{lemma:BDT}]
Let $\Gamma:[D]\times[C]\to{[D_0]}$ be a $(D^{1-\gamma},3\eps)$-disperser from \Cref{lemma:Zuc-disperser}, where $D_0^{2/\gamma}$ and $C=O(\log(D)/\log(1/\delta))=O(\log(D_0)/\log(1/\eps))$. Observe that by definition of sampler, for every $\bX$ s.t. $\hmin(\bX)\ge k$ and every $T\subseteq\bits{m}$ s.t. $\abs{T}\le \eps 2^m$,
$$\pr[x\sim\bX]{\pr[y\sim{[D_0]}]{\Samp(x,y)\in T}>2\eps}\le \delta.$$
\end{proof}
Define $\Samp'(x,y,z)=\Samp(x,\Gamma(y,z))$. We claim that for every $x$ s.t. $\pr[y\sim{[D]}]{\forall z\,\, \Samp(x,y,z)\in T}>2D^{-\gamma}$,  it is also true that $\pr[y\sim{[D_0]}]{\Samp(x,y)\in T}>2\eps$. This would imply $$\pr[x\sim\bX]{\pr[y\sim{[D]}]{\forall z\,\, \Samp(x,y,z)\in T}>2D^{1-\gamma}}\le\pr[x\sim\bX]{\pr[y\sim{[D_0]}]{\Samp(x,y)\in T}>2\eps} \le \delta,$$
 which means $\Samp'$ is a somewhere random sampler as required. To prove this, for every $x$ define $$R_x:=\{y\in[D_0]:\Samp(x,y)\in T\}.$$ Then define
 $$L_x:=\{y\in [D]: \forall z \, \Gamma(y,z)\in R_x\}.$$
 Observe that $\Gamma(L_X)\subseteq R_x$. Therefore, by definition of $\Gamma$, if $\abs{R_x}< 3\eps D_0$ then $\abs{L_x}< D^{1-\gamma}$. In other words, $\pr[y\sim{[D]}]{\forall z\,\, \Samp(x,y,z)\in T}>2D^{-\gamma}>  D^{-\gamma}$ implies $\pr[y\sim{[D_0]}]{\Samp(x,y)\in T}\ge 3\eps>2\eps$.

\section{On Random Functions and Extractors for Sumset Sources}\label{appendix:random-function}
In this section, first we show that a random function is an extractor for sumsets with low \emph{additive energy}. Similar to the size of a sumset, the additive energy is also an intensively studied property in additive combinatorics~\cite{TV06}. Then we briefly discuss why this result is not sufficient to prove that a random function is an extractor for sumset sources with \Cref{thm:affine-sumset}. 

For two sets $A,B\subseteq \bbF_2^n$, define $\gamma_{A,B}(x)=\abs{\{(a,b):a\in A,b\in B, a+b=x\}}$. Observe that if $\bA$ is the uniform distribution over $A$ and $\bB$ is the uniform distribution over $B$, then $\pr{\bA+\bB=x}=\frac{\gamma_{A,B}(x)}{\abs{A}\abs{B}}$. 
\begin{definition}
The additive energy between $A,B$ is defined as $E(A,B):=\sum_{x\in A+B} \gamma_{A,B}(x)^2$. 
\end{definition}
Without loss of generality, in the rest of this section we consider a ``flat" sumset source $\bA+\bB$ such that $\bA,\bB$ are uniform distributions over $A,B$ of size $K=2^k$. We note that $E(A,B)$ satisfies $K^2\le E(A,B)\le K^3$, and $4k-\log(E(A,B))$ is exactly the ``R\'{e}nyi entropy" of $\bA+\bB$, which is defined as $H_2(\bX)=-\log(\sum_{x\in\Supp(\bX)}\pr{\bX=x}^2)$. In the following lemma we show that if $E(A,B)$ is low (i.e. if $H_2(\bA+\bB)$ is high), then a random function is an extractor for $\bA+\bB$ with high probability. 
\begin{lemma}
For a random function $f:\bbF_2^n\to\bits{}$, $f(\bA+\bB)$ is $\eps$-close to $\bU_1$ with probability $1-2e^{-2\eps^2K^4/E(A,B)}$.
\end{lemma}
\begin{proof}
Observe that $\ex{f(\bA+\bB)}=\frac{1}{K^2}\sum_{x\in A+B}\gamma_{A,B}(x)\cdot f(x)$. Because the terms $\{\gamma_{A,B}(x)\cdot f(x)\}_{x\in A+B}$ are independent random variables, and each $\gamma_{A,B}(x)\cdot f(x)$ is in the range $[0,\gamma_{A,B}(x)]$, the lemma is directly implied by Hoeffding's inequality.
\end{proof}
Since the total number of subsets $A,B$ of size $K$ is at most $\binom{2^n}{K}^2\le 2^{2nK}$, by union bound we get the following theorem.
\begin{theorem}\label{thm:small-energy}
With probability $1-2^{-0.88nK}$, a random function is an extractor with error $\eps$ for sumset sources $\bA+\bB$ which satisfy $E(A,B)\le \frac{K^3}{n/\eps^2}$.
\end{theorem}
In other words, a random function is an extractor for flat sumset sources $\bA+\bB$ which satisfy $H_2(\bA+\bB)\ge k+\log(n/\eps^2)$. However, \Cref{thm:affine-sumset} only shows how to extract from $\bA+\bB$ when the ``max-entropy" $H_0(\bA+\bB):=\log(\abs{\Supp(\bA+\bB)})$ is close to $k$. Because $H_0(\bA+\bB)\ge H_2(\bA+\bB)$, it is possible that $H_2(\bA+\bB)\approx k$ and $H_0(\bA+\bB)\gg k$, and in this case neither of our analysis works. 

In additive combinatorics this corresponds to sets with ``large doubling" and ``large energy", and can be obtained with the following example. Suppose $A=B=V\cup R$, where $V$ is a linear subspace of dimension $k-1$, and $R$ is a random set of size $K/2$. Then $E(A,B)\ge E(V,V)\ge K^3/8$, and $\abs{A+B}\ge \abs{R+R}\approx K^2/4$.

Finally we remark that a well known result in additive combinatorics called the  `Balog-Sz\'{e}meredi-Gowers theorem"~\cite{BS,Gowers,BSG-modern} states that if $E(A,B)\ge K^3/r$ then there must exist $A'\subseteq A$, $B'\subseteq B$ of size $K/\poly(r)$ such that $\abs{A'+B'}\le \poly(r)\cdot \abs{A}$. However, if we apply this theorem on the cases which do not satisfy \Cref{thm:small-energy}, we can only guarantee that there exist small subsets $A',B'$ of size $K/\poly(n)$ which have small doubling. Because $\pr{\bA\in A' \wedge \bB\in B'}\approx 1/\poly(n)$, with \Cref{thm:affine-sumset} we can only prove that a random function is an extractor for $A',B'$ with error $1/2-1/\poly(n)$, which is comparable to a disperser.
\end{document}